\documentclass[12pt,a4paper,twoside=false,%
BCOR=12mm,%
DIV=14,%
headsepline,%
cleardoublepage=empty,%
chapterprefix,%
parskip,%
listof=totoc,%
index=totoc,%
bibliography=totoc]{scrbook}

\usepackage[ngerman, english]{babel} 

\usepackage[utf8]{inputenc}
\usepackage[T1]{fontenc}
\usepackage{mathptmx}
\usepackage[scaled=0.92]{helvet}
\usepackage{courier}
\usepackage[calc]{datetime2}
\usepackage{amsmath}

\usepackage{amsfonts}
\usepackage{amstext}
\usepackage{subfigure}
\usepackage{color}
\usepackage{graphicx}
\graphicspath{ {./fig/} }
\usepackage{ifpdf}
\usepackage{iflang}
\usepackage{makeidx}
\usepackage{nomencl}
\usepackage{setspace}
\usepackage{lscape}
\usepackage{tabularx}
\usepackage{longtable}
\usepackage{booktabs}
\usepackage{textcomp}
\usepackage{mathtools}
\usepackage[acronym]{glossaries}
\usepackage[ruled,vlined]{algorithm2e}
\usepackage[hyphens]{url}
\usepackage{pgfplots}
\pgfplotsset{compat=1.8}
\usepgfplotslibrary{statistics}
\newacronym[]{qvm}{QVM}{Qualitätsverbesserungsmittel}
\newacronym{dbas}{\mbox{D-BAS}}{Dialog-Based Argumentation System}
\newglossaryentry{decide}{name={\emph{decide}},description={}}
\newglossaryentry{we}{name={scientific institution},description={}}
\newacronym[]{gdpr}{GDPR}{European General Data Protection Regulation}
\newacronym[]{idm}{IDM}{identity management system}

\usepackage[%
    binary-units=true,
    per-mode=symbol,
    exponent-product=\cdot,
    range-units=single,
    exponent-to-prefix=true,
    zero-decimal-to-integer=true,
    list-units=single
    ]{siunitx}

\usepackage[plainpages=true,pdfpagelabels]{hyperref}
\usepackage[nameinlink,capitalise]{cleveref}
\usepackage{cite}

\def\mauve{0}
\def\graffi{1}

\newcommand{\graduationlevel}{Master}

\newcommand{\professor}{\mauve}

\newcommand{\thesistitle}{Decision Making with Argumentation Graphs}

\newcommand{\thesisauthor}{Björn Ebbinghaus}

\newcommand{\thesisauthorbirthplace}{Remscheid}

\newcommand{\thesissupervisor}{Christian Meter, M.\,Sc.}

\newcommand{\thesiskeywords}{Online Participation, Participatory Budgeting}

\newcommand{\thesissubmissionday}{30}

\newcommand{\thesissubmissionmonth}{05}

\newcommand{\thesissubmissionyear}{2019}

\IfLanguageName{ngerman}{
    \newcommand{\thesistype}{\graduationlevel{}arbeit}
}{
    \newcommand{\thesistype}{\graduationlevel{}'s Thesis}
}
\newcommand{\thesistypegerman}{\graduationlevel{}arbeit}

\ifpdf
    \definecolor{brown}{cmyk}{0, 0.81, 1, 0.60}
    \hypersetup{%
      pdftitle={\thesistitle{}},
      pdfsubject={\thesistype{}},
      pdfauthor={\thesisauthor{}},
      pdfkeywords={\thesiskeywords{}},
      colorlinks=false, urlcolor=blue, citecolor=brown,
      bookmarksnumbered=true,
    }
\else   
\fi

\makeindex

\begin{document}
\onehalfspacing

\frontmatter

\pdfbookmark[0]{\IfLanguageName{ngerman}{Titelseite}{Front Page}}{frontpage}
\begin{titlepage}
  \centering
  \includegraphics[width=5cm]{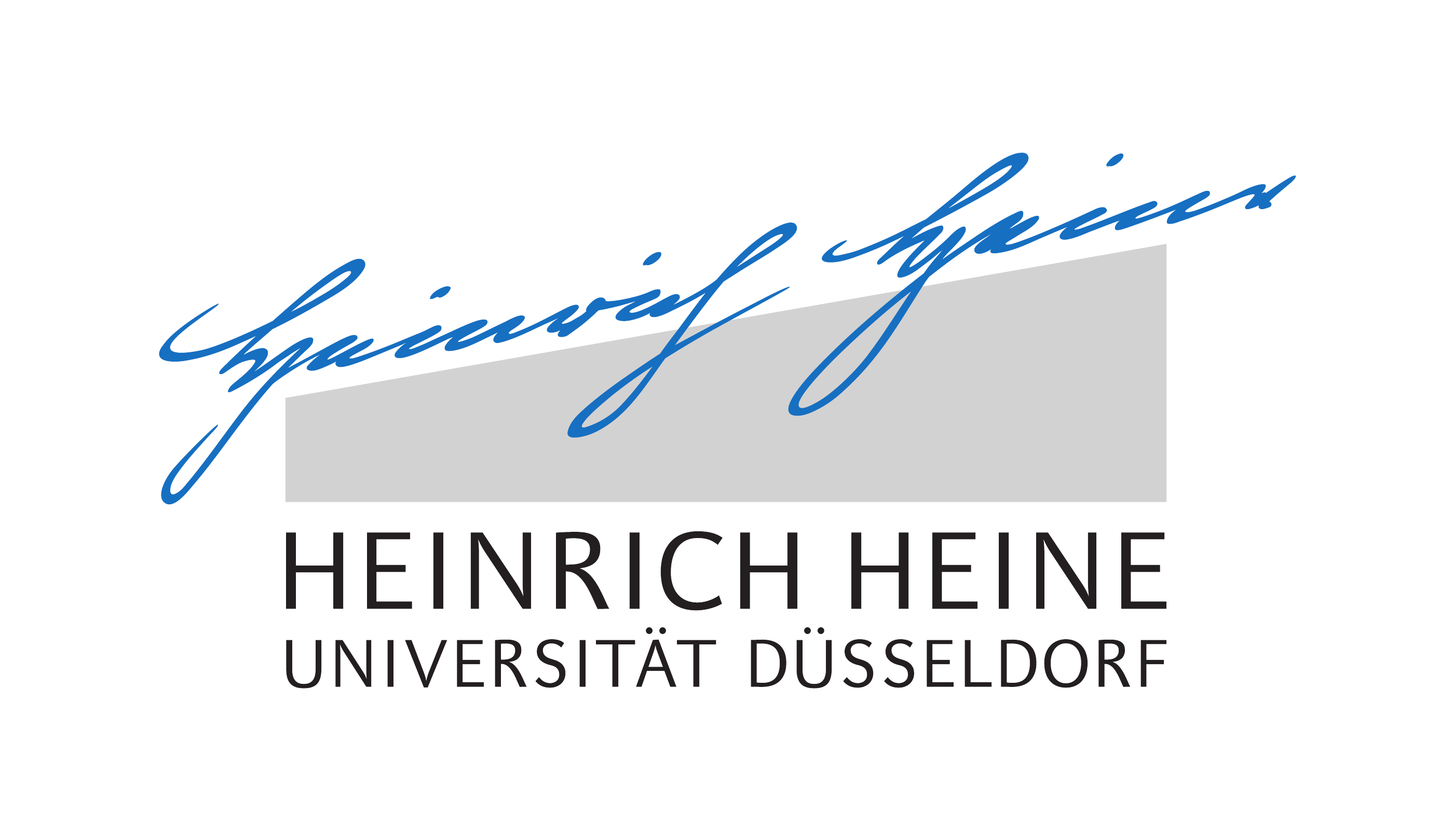}\\

  \vfill
  \huge
  \thesistitle{}\\*[40pt]
  \normalsize

  \vfill
  \large
  \thesistype{}\\[0.25em]
  \normalsize
  \IfLanguageName{ngerman}{von}{by}\\
  \Large
  \thesisauthor{}\\

  \vspace{5mm}
  \normalsize
  \IfLanguageName{ngerman}{aus}{born in}\\
  \thesisauthorbirthplace{}\\[1cm]

 \if\professor\mauve
    \IfLanguageName{ngerman}{vorgelegt am}{submitted to}\\[5mm]
    \IfLanguageName{ngerman}{Lehrstuhl für Rechnernetze}{Professorship for Computer Networks}\\
    Prof.\ Dr.\ Martin Mauve\\
  \else
    \if\professor\graffi
      \IfLanguageName{ngerman}{vorgelegt bei}{submitted to}\\[5mm]
      \IfLanguageName{ngerman}{Technik Sozialer Netzwerke}{Technology of Social Networks Lab}\\
      Jun.-Prof.\ Dr.-Ing.\ Kalman Graffi\\
    \fi
  \fi

  \IfLanguageName{ngerman}{Heinrich-Heine-Universität Düsseldorf}{Heinrich-Heine-University Düsseldorf}\\[0.5cm]
  \DTMmonthname{\thesissubmissionmonth} \thesissubmissionyear{}\\[0.5cm]
  \IfLanguageName{ngerman}{Betreuer}{Supervisor}:\\
  \thesissupervisor{}
\end{titlepage}

\cleardoublepage

\pdfbookmark[0]{\IfLanguageName{ngerman}{Zusammenfassung}{Abstract}}{abstract}
\begin{center} 
    \huge \IfLanguageName{ngerman}{Zusammenfassung}{Abstract}
\end{center}


This work is about making decisions by digital means.
Funds should be distributed by the students of Heinrich-Heine-University.
The proposals were made by the students themselves without further influence.
For this purpose, dialog-based argumentation is used to give the participants a better understanding of various arguments.
In addition, a software service has been developed which allows the students to express their preferences for various proposals.
An experiment was carried out at the university, which should prove whether students are satisfied with this type of participation.

The results indicate that the procedure is well accepted and thus successful.
However, improvements to the process itself were necessary during the experiment and should be considered for future procedures.
Further procedures are desired.

\cleardoublepage

\pdfbookmark[0]{\IfLanguageName{ngerman}{Danksagung}{Acknowledgments}}{acknowledgments}
\begin{center} 
    \huge \IfLanguageName{ngerman}{Danksagung}{Acknowledgments}
\end{center}


Huge thanks to Christina and Markus, for proof-reading this wörk and keeping me hydrated. 
I developed a habit of bringing water to the office now. 

I am thankful for the guidance Christian provided, as I could ask any time for advice when I got stuck with a specific problem.
This applies both to this work and to the time before it.

Also, I am grateful towards our weekly Tuesday jour fixe.
Having a place where I can express problems and getting profound comments and suggestions is a blessing. This provided me with new food for thought.

\cleardoublepage

\pdfbookmark[0]{\contentsname}{content}
\tableofcontents

\listoffigures

\listoftables

\mainmatter

\cleardoublepage


\chapter{Introduction}
With the rise of the internet, people are able to communicate with each other in near real-time. 
This enables us to connect people to enhance their life in a meaningful way.
Communication is key to this kind of society.

This work will explore the possibility to make decisions for budget allocations with the support of dialog-based argumentation. 
It extends the \gls{dbas} with a way to assign costs to proposals and provides a separate system for information and voting on said proposals.

\section{Participatory Budgeting}
Participatory budgeting is a decision process in which common citizens are involved to allocate funds. 
The ruling over these funds was mainly in the hand of the acting government and was done behind closed doors, often with none or just a little information for the public to comprehend. 

With the desire, and in some cases even need, for transparency a government is concerned to involve its citizens into the process. 
The first step is providing information about the internal processes of the administration. 
Several open data initiatives, with the goal to make this kind of information accessible to the public, began to grow and are even getting implemented as law\cite{open_data}.

The next step is to involve citizens actively, by allowing some kind of influence in the process.
This can reach from surveys, which are used as advice
by councils to make better decisions for the public, and can go so far as to leave the entire decision to the citizens, with former decision-makers just taking a steering role in the process.

\section{Goals of this Work}
The goal of this work is to research whether it is possible with a rather small group of students from the computer science department, to have a deliberative process of rational proposal making, educational argumentation, and a direct voting procedure.

This work provides the means to enable the participants to take part in the participatory budgeting process by expressing their most important positions in the discussion.
It has been made possible for the participants to prioritise their arguments in the service called \gls{decide} against each other after they had been discussing them between themselves in \gls{dbas}.
The priorities are then used to rank proposals against each other to form a fitting result set, which stays inside the budget and is simple enough to be understood without extensive knowledge in computational social choice.

It should be observed whether the ranking by a specific method generates the best sense of fairness for the participating users.
An understandable and accepted result is desired. 
For a better acceptance of the result, a brief explanation of how the decision is made was given.

An experiment should be conducted to test if there is interest in this form of participation. 
The experiment is not artificial, but is a real process, with real money, which is scientifically accompanied.
It, therefore, is as close to reality as it gets with the hope to introduce this kind of participation to more areas, like other departments. 
It should be used to improve the procedure iteratively.

\section{Structure}
The work will begin with an introduction to established and proposed ways to argue online.
This will show the advantages of argumentation systems over other means of participation.

Thereafter, a short introduction to voting and scoring systems will be given.
It will just scratch on the surface of computational social choice but gives enough detail to understand the used scoring system and an outlook what would (computational) be possible.

Then the modifications to \gls{dbas} and the features for \gls{decide} are presented. 
It is explained why certain decisions were made to satisfy the needs of the experiment.
Also, it contains an example that was given to the participants for them to understand the voting procedure.

Afterwards, the execution and the results of the experiment will be discussed.
This includes a part of the survey, which was done in cooperation with the sociologists of the Heinrich-Heine-University.
In addition, it contains suggestions for improvements to the systems used, \gls{dbas} and \gls{decide}, which should be considered for future procedures.

To compare this work to other methods, three different participation procedures will be presented. These range from completely analogue to digital methods not unlike the procedure used here. 
For two of them, the differences are listed in detail, as they are similar enough to this work, that ideas could be adapted.

Finally, a summary of the lessons learned will be given, along with general considerations of where the topic might go in the future, and what should be explored to compare this work.


\chapter{Argumentation Systems}
This chapter gives an introduction to argumentation systems and highlights their advantages versus traditional ways of expressing opinions online, like forums or comment sections.
It assumes, that more advanced systems allow for a more defined way of information storage, which can be used to confront the participant with different aspects of a topic in a more efficient way.

\section{Motivation}
For participants to engage in a decision-making process, there has to be some kind of understanding of the problem.
While each participant could inform herself about a disputed topic to form her opinion, this method of opinion-formation will result in narrowed down view on the problem at hand.
Getting knowledge about the reasoning of other participants can prove complicated, as we live in a world where (nearly) everyone can communicate in real-time.
Everyone can submit their own statements and receive the opinions and arguments of each other via means of the internet, but they are often unstructured, like in forums.

\section{Forums}
Forums are one of the oldest forms of online participation\cite{lueg2012usenet}, where users can argue about a topic.
The problem with argumentation in a traditional forum is, that the argumentation gets harder or even impossible with a growing number of participants. 

\begin{figure}
    \centering
  \begin{minipage}[t]{0.49\textwidth}
    \includegraphics[width=\textwidth]{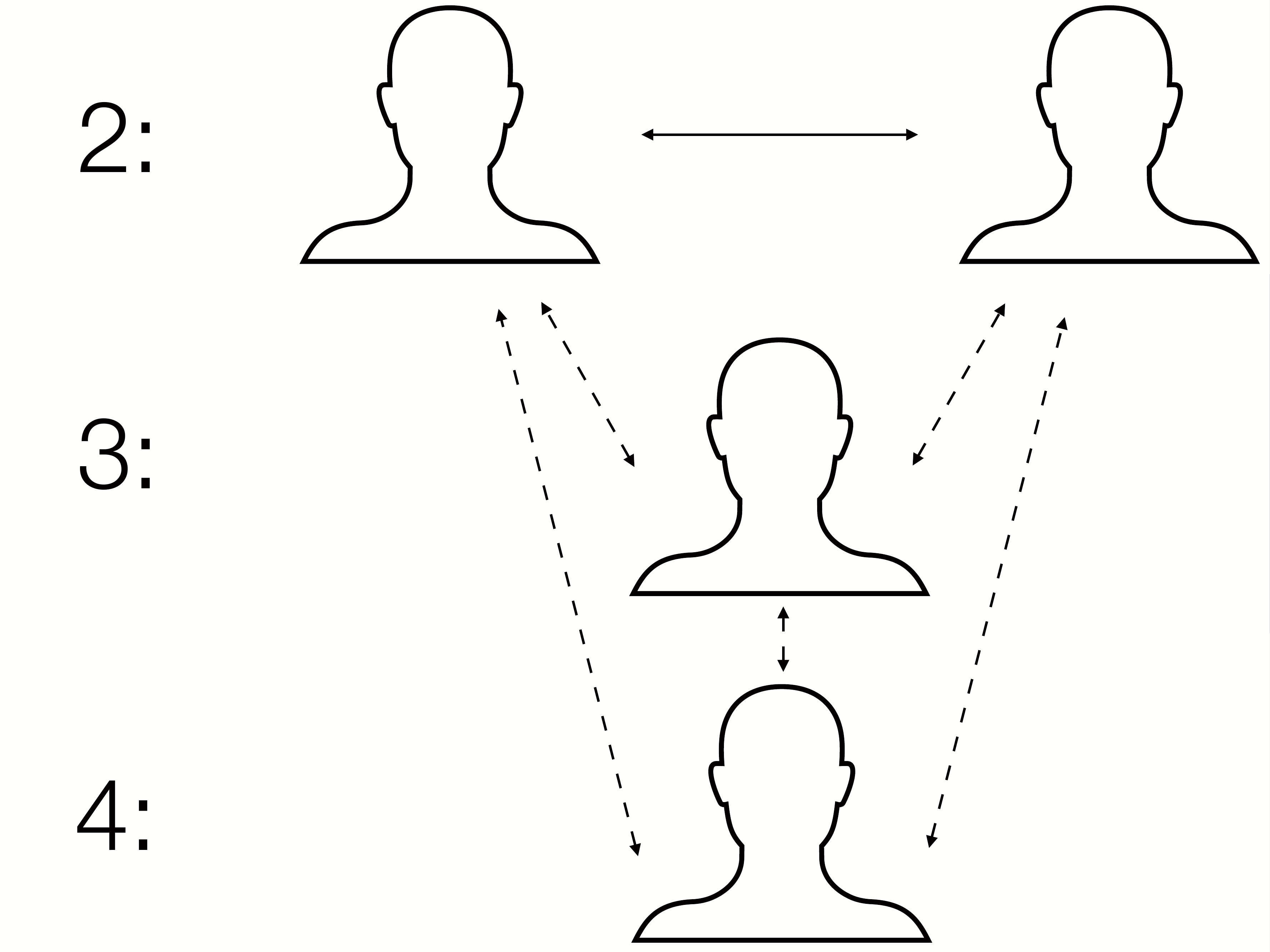}
    \caption[Exploding complexity with a growing number of users]{Exploding complexity with a growing number of users, as each user has to argue with \(n-1\) other users.}
    \label{fig:forum_complexity}
  \end{minipage}
  \hfill
  \begin{minipage}[t]{0.49\textwidth}
    \includegraphics[width=\textwidth]{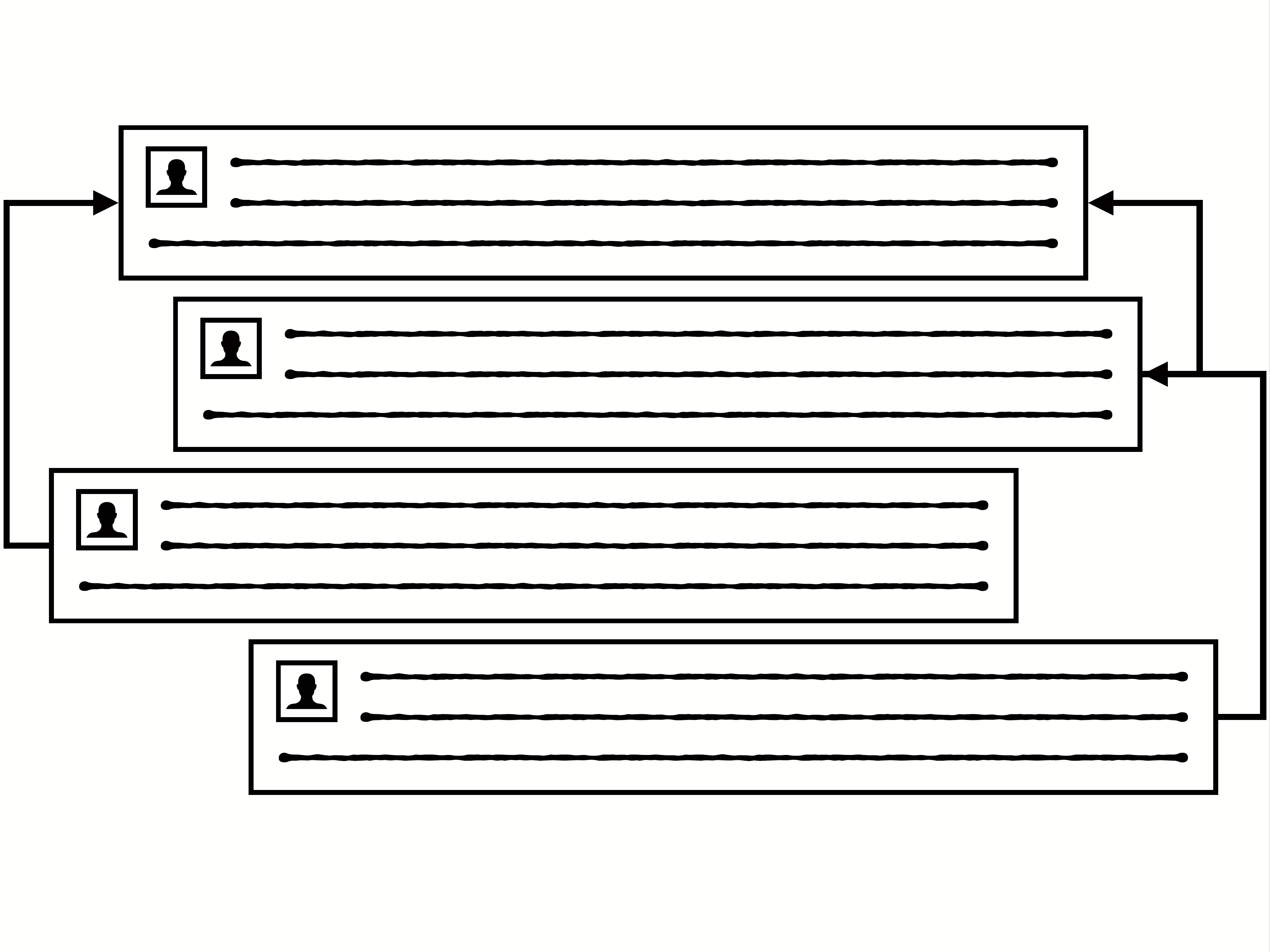}
    \caption[References inside a forum thread]{References inside a forum thread. This leads to a confusing and not comprehensible argumentation, because they are perceived in the order of creation.}
    \label{fig:forum_order}
  \end{minipage}
\end{figure}

Imagine an argumentation between two participants, that goes back and forth. 
If a third participant enters the argumentation, this can get quite irritating, as the new user would probably have another mindset then the other participants, influencing the argumentation in a way, that disturbs the original argumentation structure.
To maintain the overview over where the thread is going with its topic the participants have to make sure that they read each message from each other participant like shown in \autoref{fig:forum_complexity}.

You can often see posts in forums, that refer to a post several posts before their own, addressing the author of this post directly by name to indicate their interest of continuing the argumentation at this point like shown in \autoref{fig:forum_order}.
This creates a sub-thread in a forum thread, that is not supported by the linear data structure of the forum.

Another disadvantage is that this form of participation, through the expression, of thoughts results in large, complex texts and are more comparable to a letter exchange as they are to a real-time argumentation. 
Made statements are not just confined to a single thought, as users often submit them in a length of several sentences to full articles.

\section{Comment Sections}\label{sec:comment-section}
Nowadays, in times when every kind of information is available at all times, forums made way for a more rapid type of expression, in form of smaller snippets, which are done through comments known from pages like Facebook, Reddit or even Twitter.
These systems allow for a restricted way of branching out from different statements a participant has done.
All three examples allow for nested comments, where a participant can react to a previous statement later on, without disrupting the current argumentation, in a branch. 
Although the comments can be nested, there is often a limit on how deep an argumentation can go, e.g. Facebook allows only for two layers, making a compromise between functionality and clarity. 

The greatest drawback of these systems is that they are missing any information about the attitude of the participants. 
Participants are not encouraged to add separate information, on how their argument is to understand. 
Entries in this systems lack semantic information, as a user can not enter the attitude of the entry, like if the entry is a counter-argument against the entry before.
Or if the entry is a new piece of argumentation, not tied to a preceding argument.

\section{Argumentation Systems}
Argumentation systems are trying to fix the aforementioned problems by guiding the participant to establish a context in an already existing argumentation for her own arguments. 
They allow modelling of relations between (at best) atomic statements, so understanding the context, in which a statement has been made, becomes simple as a this-because-of-that argument.
The statements and arguments are therefore representing a graph which can grow indefinite in depth and width. 

The edges in this graph can contain information, which is missing in comment sections. 
A participant can add his statement with the information, that it is an argument against another statement from another participant. 
This results in a system where statements from users are much more tied to their intended meaning, giving them more semantic meaning which can be understood and used by machines.

Although there are ways to structure the content that is contributed, the amount of content which is gained from hundreds or even thousands of participants can be overwhelming at times.
Each participant would have to asses present arguments and sort their statements correctly into the right position, adding the necessary data to build the graph, which can be quite inconvenient, especially for inexperienced users.
If these relations in the graph are not utilized, it will result in the practical loss of data, as a seemingly new argument will be added, while the same argument may have been made before results in a partition of arguments.
This reduces the diversity of arguments and introduces redundancy.
Merging this grown apart parts of the argumentation later on will be quite laborious.
This is not a problem introduced by argumentation systems, this is present in the earlier systems as well, although an argumentation graph can be utilized to deal with this.

\section{D-BAS --- The Dialog-Based Argumentation System}
\begin{figure}[h]
    \centering
    \includegraphics[width=.5\textwidth]{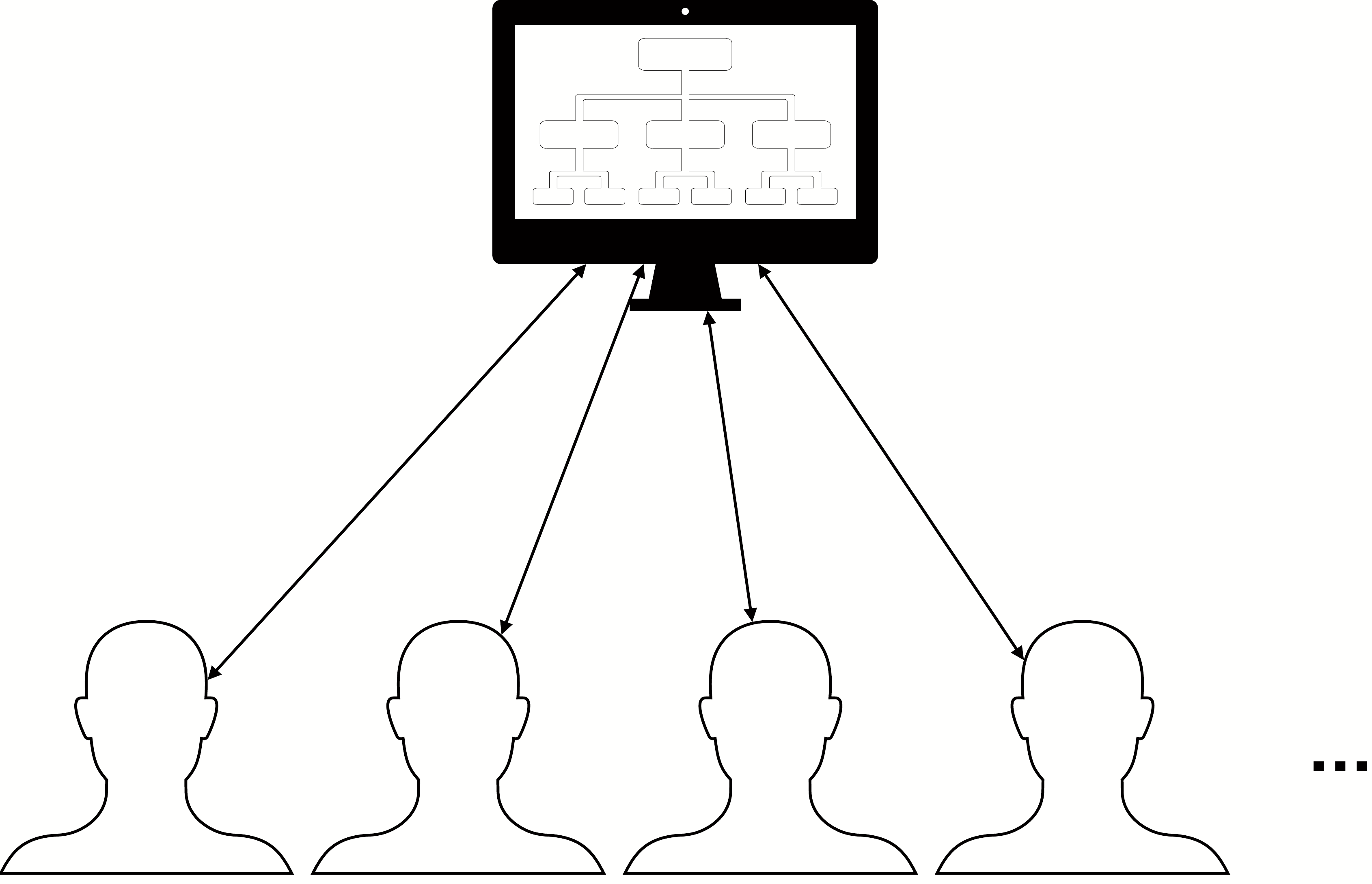}
    \caption[Horizontal scaling of argumentation via an automated intermediary.]{Horizontal scaling of argumentation with via an automated intermediary. Each user just has to argue with the system.}
    \label{fig:horizontal-scaling}
\end{figure}

\gls{dbas}\footnote{https://dbas.cs.uni-duesseldorf.de}\cite{krauthoff2016a}\cite{krauthoff2018d} is an argumentation system which makes use of the argumentation graph to allow participants to take part in an argumentation, naturally by letting them interact via a dialog like interface.
This allows for a scoped interaction, where the participants are given just a partial view on the argumentation, reducing the perceived complexity of the argumentation graph. 
Having the system itself as one's argumentation partner enables horizontal scaling of the argumentation. 
Users do not interact directly between each other, this can be seen in \autoref{fig:horizontal-scaling}

\subsection{The Argumentation Graph}
The argumentation graph of D-BAS is the structure in which arguments are stored. 
The graph is acyclic and directed, though the acyclic property is not enforced by the data structure and could be utilised, later on.
These directional circles are not possible to create with the means of the user interface and are therefore not considered in the rules of the dialog game execution platform.

The dialog game execution platform is the logic behind \gls{dbas}, that chooses on how to proceed in an argumentation by consulting the given argumentation graph and the user's current location in it.

\begin{description}
    \item[Issue] 
    At the root of every argumentation graph there is an issue node, containing all information about the current topic, like the title, a summary, flags whether this discussion should be writeable etc. 
    An issue is the scope of the discussion a participant wants to take part in and it gives context to all statements in the discussion. 

    A decision process, like it will be presented in this work, would be represented by a single issue. 
    
    \item[Position]\label{dbas:position}
    Positions are the nodes that are connected to an issue node. 
    They are sub-topics or proposals on the topic that a user can make. 
    The users then have the option to agree or disagree on a position starting their path through the attached graph. 
    Positions cannot be reused in the argumentation as they are often neutral in their meaning, not suited to support or attack another argument.
    This is by design and not by the necessity of the argumentation graph. 
    For example, systems like EDEN\cite{MeterSchneider2018a}, which are developed to interconnect \gls{dbas} argument graphs, have no special role for positions.

    Positions are a special case of Statements.
    
    \item[Statement] 
    Statements are the reusable part of the graph. 
    They are intended to be atomic, meaning they just contain their meaning themselves.
    Every context in which they would be understood differently is provided by their usage inside the graph.
    For Statements to be able to exist in the graph they need to be established in a context in the form of a premise.

    Statements are not confined to a single issue, as one goal of \gls{dbas} is to reduce redundancies.
    This allows the user to reuse already made arguments, connecting the sub-graph of another issue to the current one.
    
    \item[Argument] 
    The graph has to be connected, of course, this is the job of arguments. 
    Arguments are directed edges. 
    Their root is one or more premises and they connect them to another statement or when making an undercut, even another argument. 
    More on undercuts in the next section.

    The endpoints of the edge are called conclusion.
    The arguments contain the information whether a premise is used to attack or to support the conclusion, this is called the attitude of the argument.
    Arguments are either an attack or a support.
\end{description}

\subsection{Relations in the Graph}
To use the graph a hand full of higher relations between nodes are defined, with which the participant will traverse through the graph. 
They relate to proper human argumentation techniques and are made directly available through user interface options. 

\begin{description}
    \item[Support] 
    Altogether a support is every argument with a positive attitude like every negative argument is called an attack, but since there is just one relation with this property this term is used twice.

    An argument \(A\) supports an argument \(B\) when the argument \(A\) has a positive attitude towards the premise of the argument \(B\). 
    A support can be rebutted on the conclusion of argument \(A\), through an \emph{undermine} of \(B\).

    \item[Rebut] 
    An argument rebuts another argument when the arguments are of opposing attitudes and share the same conclusion. Example: `\(A\) is a good reason for \(C\), \emph{but} there is proposal \(B\), which is a reason against \(C\).'

    \item[Undermine] 
    An \emph{undermine} is an attack on the premise of another argument and therefore the opposite of a support. 
    An attacking argument has a negative attitude towards the premise of the attacked argument. 
 
    \item[Undercut] 
    An argument is an \emph{undercut} when its attitude is negative towards another argument. 
    This occurs, when the participant has the opinion, that the premise of the attacked argument is a correct statement, but the usage of it in the argument is invalid. 
    The premise and the conclusion do not relate together.
\end{description}

These relations are not complete, as they do not mention for example an argument, which has a positive attitude towards another argument.
This was omitted as they are too complicated to comprehend, and were rarely used anyway.
Experiments have shown, that the majority of arguments, which do not represent positions, are attacks\cite{Krauthoff2017a}.
This is probably a result of D-BAS trying to provide the user with a counter-argument, which therefore always forces the user into a defensive role.

\begin{figure}[h]
  \centering
  \includegraphics[width=\textwidth]{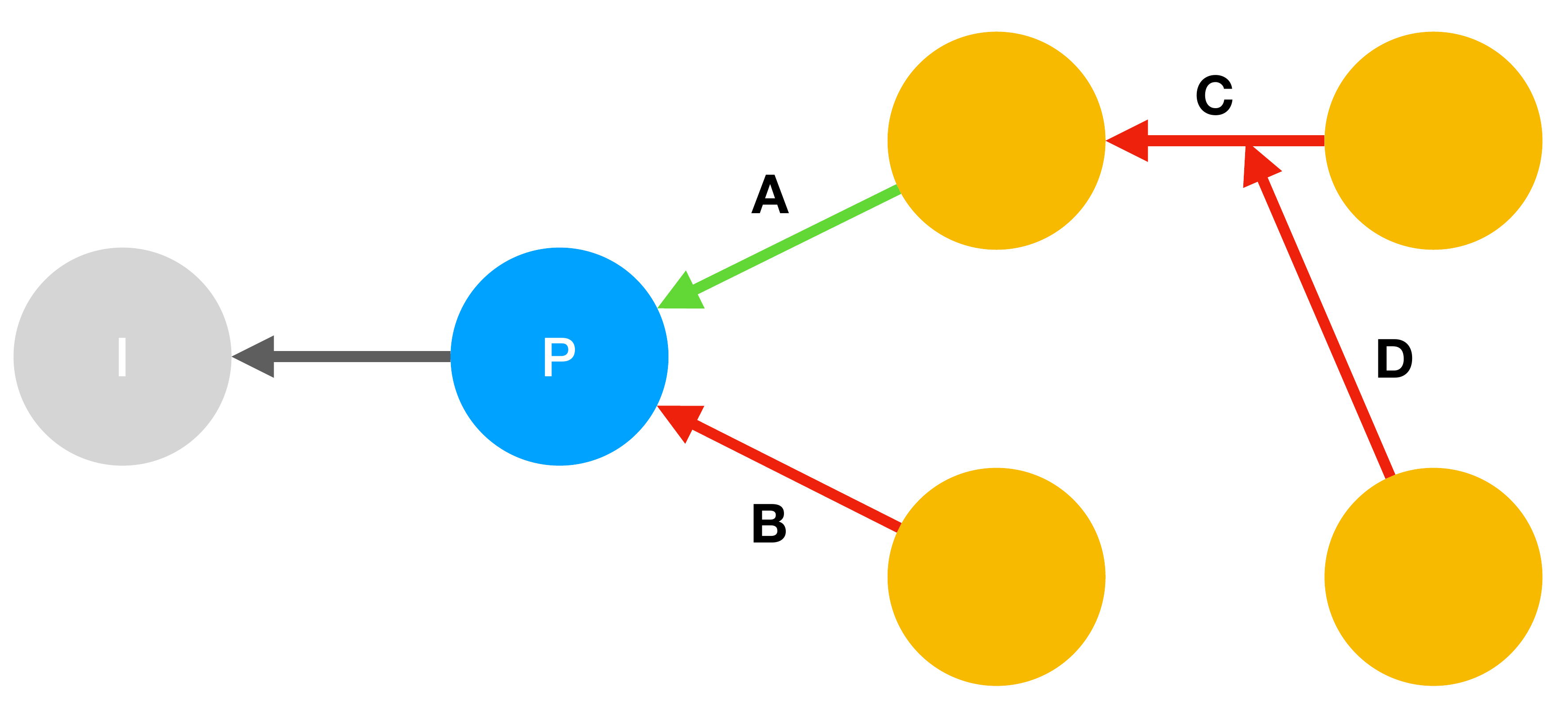}
  \caption[Example of the argumentation graph.]
  {
    Example of an argumentation graph.
    I denotes an issue, P a position, A-D are arguments. The targets of the edges are called conclusion, the sources premise.
    \\ A supports P.
    \\ B rebuts A on P, by attacking the conclusion of the supported P.
    \\ C undermines A, by attacking its premise.
    \\ D undercuts C, by directly attacking C.
  }\label{fig:arggraph}
\end{figure}

\subsection{Moderation of the Argumentation}
\Gls{dbas} has a distributed moderation system, where the users are empowered to vote for or against changes of the argumentation graph.
Statements that should be reworked can be reported by the users and are shown in queues, where it can be voted on these changes. 
Changes are approved, when the first side ---pro or against--- reaches five votes, or one side has three votes more than the other side.

Possible changes include the deletion of inappropriate statements, correction of spelling errors or the mere proposal, that a statement should be edited, but without a new proposal.
Furthermore, there are more advanced moderation features, such as splitting an argument in two or merging two same arguments to one.


\chapter{Voting and Scoring Systems}\label{chp:voting-and-scoring}
This chapter shows up some methods a vote can take place and how these votes can be scored to deliver a result that fits.
It will be confined to what is necessary for applying them and their basic advantages and disadvantages, not going in too deep into the field of computational social choice.
Priority is that the procedures stay understandable to an untrained participant.

\section{Motivation}
To make a decision there has to be a foundation on how to decide. 
It needs rules on how a certain decision will be made or not. 
Of importance is the difference between how a decision is made by an individual and a decision that will be made by a group. 

A human being by itself compares a set of choices and weights them against each other by certain criteria, such as personal implications, experiences and own goals.
These criteria may work for the decision of a single person but are getting complex when there is a group of people. 
Even for two people, the challenge is to align their needs and desires, even though they can communicate efficiently.
In a larger group of people the possibility to communicate, and as a result negotiate, is vastly reduced.
Thus making it impossible for anyone to decide in a way that fits best for all.

We have votes to tackle this problem.
Voting is nothing other than the aggregation of several personal decisions, namely who or what is best for themselves and their relatable environments, like your family and friends.
Several ways on how to vote on something and how to aggregate these votes for a final decision exist.
Different forms of voting and scoring have different results and because of this, they are suitable for different problems. 
Voting procedures depend on what is voted on. 
Be it the vote for the single next leader of a group or, like in the case of this work, the best way to distribute money to different proposals in a way most voters will be satisfied with the what is going to be implemented.

Fact is that most voting processes are kept simple, to allow the voters to comprehend how the decision is made. 
This provides confidence in the process and leads to more acceptance for the result.
If the process seems fair to everyone, then everyone should agree with the outcome, even if the outcome is not in their favour.

\section{Approval Voting}\label{sec:approval}
This form of voting is one of the simplest voting systems.
The voter is given a binary choice to either approve or reject a candidate --- giving a score of 1 or 0.
All candidates scores are then added together. 
The winner of the procedure is the candidate with the highest sum of scores.

An advantage of this scoring system is that voters are able to vote for whomever candidate they wish. 
They can, if they like, vote for every candidate they approve to win.
Because of this, a voter can cast a vote without the fear of throwing this vote away for a candidate they like but has low chances of winning.
By doing this, they can actually unwillingly support an unwanted candidate, as they deny their vote to an actual rival of this unwanted candidate.
Such a voting behaviour would split the votes between two or more candidates, resulting in a smaller proportion of all the votes for each candidate.

Imagine a scenario with four candidates A, B, C and D.
The majority of two-thirds of the voters would like to see A, B or C to win, and would be satisfied enough if anyone else wins except D. 
They, therefore, spread their single vote over A, B and C, giving each of them 2/9th of the total votes. 
The third of voters who want D to win vote just for D, giving it 1/3rds (3/9th) of the vote.
D wins the vote even though the majority of 6/9 does not vote for this.

Furthermore, the voting procedure is relatively simple, so that no deep knowledge is required from the voters and the procedure takes place with regular ballots. 
Like stated in the section before, it is preferable to use a voting procedure that is easy to understand, as this enables the voter to have trust in the result.

A drawback of this kind of scoring is that it encourages bullet voting. 
This is a form of strategic voting, where a voter votes just for her single most favourite candidate with the intention of not providing any points to the alternatives. 
This is possible, because voting for a candidate which is acceptable, but not the most preferred, creates competition for that most preferred candidate.

A noteworthy property of this voting system, in the light of this work, is that the system allows adding a candidate later without the need that the voters have to cast their votes anew.
The new candidate does not change the result if no one votes for it.

\begin{figure}
  \centering
  \includegraphics[width=.58\textwidth]{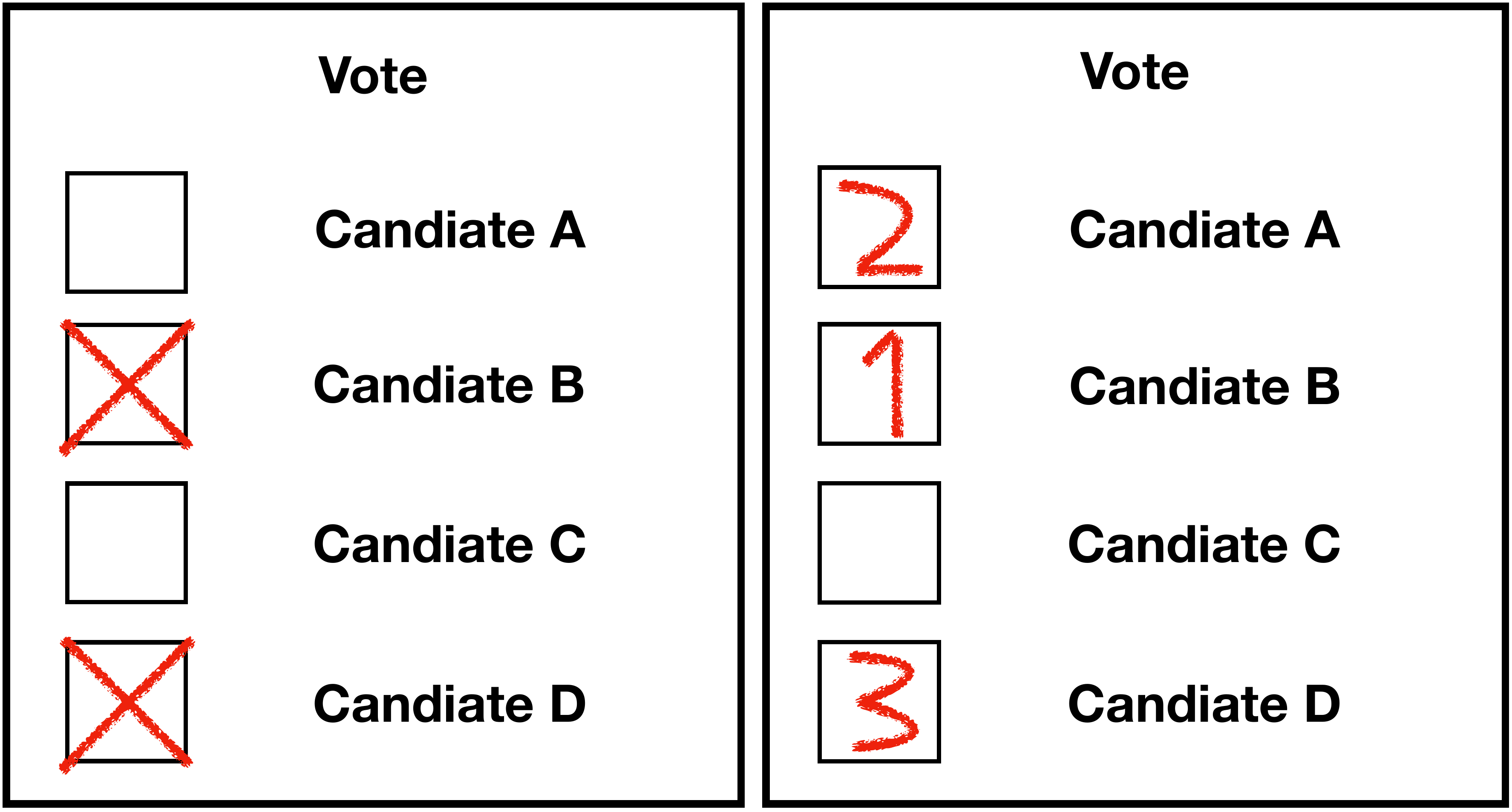}
  \caption[Ballots for the voting methods.]{The ballots for the voting methods. Left: Approval Voting, Right: Borda Count}
  \label{fig:ballots}
\end{figure}

\section{Borda Count}\label{sec:borda}
Borda count is a scoring count that asks the voters to rank the choices from most favourite to least favourite, which is, therefore, a preferential voting system. 
It requires that the amount of choices is not changing and the voter casts her vote for every choice.

A total of \(n\) choices are scored, so that for each voter, her choices \(c_1, c_2, \dotsc,c_{n-1}, c_n\) are denoted the scores \(n, n-1, \dotsc, 2, 1\), which are subsequently summed up for a final score.
No tiebreaker is included in the traditional voting. 

An alternative is that instead of starting with \(n\) as the highest and ending with \(1\) as the lowest score, \(n-1, n-2, \dotsc, 0\) is used as the scores\cite{Black1987}.
This results in a different distribution of relative scores, for example in a vote with two choices the first vote will get all the points, while the second one gets none.
In the traditional Borda count, the first vote would get 66\% of the score (2), while the second one is still getting 33\% (1).

Another system is the so-called Dowdall system used in the voting process of Nauru\cite{reilly2002social,borda_Nauru_Slovenia}, where the scores are not affected by the total amount of choices, but their position alone.
The scores are distributed by the harmonic progression, so they result in \(\{c_1\rightarrow1, c_2\rightarrow\frac{1}{2}, c_3\rightarrow\frac{1}{3}, \dotsc, c_n\rightarrow\frac{1}{n}\}\).
This, of course, strengthens the vote for the first preference and lessens the one for the later preferences.

\subsubsection{Partial Borda Count}
As it is often not feasible to let a voter rank all possible choices, modifications to the traditional Borda count have been evolved. 
Most of the practical appearances of some sort of partial Borda counting choose to set a rule on how many votes a voter has to cast. 
For example, Papua New Guinea requires the voter to cast at least three votes\cite{Emerson2013}.

All of these scoring systems with partial votes have to deal with the problem on how to score the choices that were not chosen by a voter.
The simplest solution is to assign the minimal value to them (e.g. 0 or 1). 
Also, there are proposals to use the average score the remaining choices would have got if they were chosen\cite{dummett1997principles}. 

\subsubsection{Borda count with multiple winners}
All presented methods till now consider just a single possible choice as the winner.
Since in a scenario where an unknown number of proposals fit in a given budget, it is necessary to conduct a vote with more than one winner, or even an unknown number of winners.
Though in most cases the number of winners is known beforehand, for example in the vote for seats in a parliament. 
One way to manage this problem is to just declare the choices with the highest Borda scores the winners. 
Other solutions like Quota Borda Count\cite{dummett1984voting} exist, but they are not used.

\section{Score Voting}
With the so-called score voting the participant is required to not just approve a candidate, but to score them on a given scale.
This could be a numerical point value, like a scale from one to ten, or a semantic value.
An example for that may be \emph{Approve}, \emph{Disprove} and \emph{No Opinion}, where the options can be related to a value of \(+1\), \(-1\) and \(0\) respectively.
A scoring vote with two options is equal to the approval voting.

This can be helpful as it allows for more precise control over one's voice.
Candidates who are more useful would get a better ranking than less useful, but still desired, candidates.
Image the case that a participant has to rank two candidates.
There would be no way to express that the participant prefers both candidates the same. 
The other way round it would be impossible to have a candidate way more insignificant to the final decision.

\subsection{Direct Resource Allocation}
For a resource distribution decision like present in this work, one could assume that it would be wise to use score voting, with the modification that one can not score a candidate on an independent scale, but has to distribute a fixed number of points over the candidates.
Euro could be the resource used in the procedure.

This may sound like a good idea in the first place, because it could give a feeling of direct control over the funds to the participant, as they can directly allocate the funds, without any intermediate steps given.
This is a bad idea since it would not guarantee that any candidate would get the needed funding to even be viable for implementation.

\subsection{Decisions over Proposals Based on Their Utility}\label{subsec:utility}
Scoring with a numerical value could be used to gather the utility from a participant, meaning how high the perceived importance of a proposal is for a participant.
This indicates the impact on satisfaction with the result. 
With this information, it would be possible to use algorithms, that aim to optimize the average satisfaction of the participants. 

It was decided to not use this kind of scoring, as a goal was to keep the process as simple as possible, in the hope to keep transparency and trust high in the procedure.
Using a complex method with several input weights would result in a procedure which would not be easy to understand by an untrained participant.

\begin{description}
  \item[Maximize average satisfaction] \hfill \\ 
  This will result in a great outcome on paper. A higher grade of satisfaction of the voters has the appearance of a great voting result. 
  But it opens to the possibility that an algorithm declares a minority the winner, just because their satisfaction would balance out the dissatisfaction of the majority.

  \item[Maximize median satisfaction] \hfill \\ 
  Maximizing the median of satisfaction could neglect minorities. 
  The majority could experience great satisfaction with the result, outweighing the dissatisfaction of the losers.

  \item[Minimize dissatisfaction] \hfill \\
  This optimization goal removes the problem of a high loss for the losers, which would allow for low satisfaction altogether. No one would be overly dissatisfied, but on the other hand, eventually no one will gain much.
  Everyone could be a loser with this as a strict goal.

  \item[Make the algorithm as transparent as possible] \hfill \\
  The goal chosen in this work is to make the deciding algorithm as transparent as possible. 
  This is based on the thought that satisfaction may not just be dependent on the result itself, but the process which leads to it. 
\end{description}


\section{Making a Decision with Multiple Winners}
In a procedure where there is more than one winner, the set of winners has to be selected in a way suited to the kind of decision which is made.
Obviously, in a budget decision scenario, there is often the case that a lot of the top scoring candidates actually cannot win because of budget constraints. 
In this case, it has to be decided on what to aim for in the decision. 
Should a single, costly proposal that uses up the whole budget be able to win on its own?
Or should instead the next, possibly cheaper proposals win, as this would allow for less of a monopolization of the budget?

In the experiment done for this work, it was opted to go the route of the simplest decision algorithm, as it was aimed for transparency.
For example, it could not be allowed that the proposal with the highest score is not in the winning set.
While this would be possible with some algorithms, it would seem odd for the participants.

\section{Fairness}
Consider the following scenario: There are three proposals \(A\), \(B\) and \(C\). 
All three proposals have the same cost, but the budget is just enough for a maximum of two. 
A vote with a hundred participants is held for these proposals and fifty voters voted for \(A\) and \(B\) and fifty voted for \(C\). 
Everyone, who voted for \(B\) also voted for \(A\), and no one who has voted for \(C\) has voted for \(A\) or \(B\).
The proposals are scored, which results in \(A\) getting the highest score, \(B\) a bit less than \(A\) and \(C\) a bit less than \(B\). 
Which possible set should be selected as the winning set? 

An algorithm can dictate that \(A\) and \(B\) have to win, as they have the highest score. \(C\) would not be included in the winning set, because it can not fit into the budget. 
The decision is easy to understand for every participant.
Should it be expected that everyone will be satisfied with this decision because of this?
Imagine a voter who voted for \(C\). 
This voter did not gain anything from the budget, while she sees another voter for \(A\) and \(B\) who wins it all.

One could argue that the winning set of \(\{A, C\}\) would be fairer, as now everyone wins at least something. 
Sure, voters of \(\{A, B\}\) would have a worse result, but at least half of their favourites win. 
Both voting groups would have their most favourite candidate included in the winning set.
This could lead to satisfaction with the result for everyone, even if the idea that some candidate could lose with a higher score sounds counterintuitive.
By making the decision this way, it would profit from a utility factor as mentioned in \cref{subsec:utility}, as it would be possible to assess whether losing a winner is acceptable for the voters of \(\{A, B\}\), and it could, therefore, be weighted of if a compromise can be done without appearing unfair. 
It could improve the situation for everyone.

\chapter{Decision Making with D-BAS}
This chapter shows how the decision process was designed.
The goal was to make use of the educational advantages of \gls{dbas} for budgeting decisions. 
Design decisions were made with the fact in mind, that the possible participants in this procedure would not have any experience with argumentation systems and participatory budgeting methods.

\section{Motivation}
Reaching a large group of people and allowing them to each participant in a decision with equal opportunities is a challenging task which clearly prefers the use of online tools. 
As the goals of \gls{dbas} are to provide an educational argumentation that scales and stays objectively, even if the arguments are getting specific, this system appears like a great choice for an improved self-consultation process. 

Since these procedures vary from situation to situation, it must be possible to adapt the procedure accordingly. 
While in early stages of the design, the aim was to utilize the \gls{dbas} argumentation graph, but not necessarily the \gls{dbas} interface itself, it turned out to be better to use the already proved interface as well to avoid a scope creep which could threaten a live usage.
A modification of the argumentation flow itself could nevertheless lead to interesting ways of interacting with the participant and may even allow for advanced decision-making techniques like negotiation.

\section{Adding of New Proposals}\label{sec:add-new-proposals}
In the terms of \gls{dbas} a proposal would be a position. 
Positions can be argued about and the first step after selecting a position is to agree or disagree on them, which is perfect for a scenario like decision-making.

For positions to accommodate for the costs, they had to be modified.
The modification added a plain cost field that every participant had to fill in when they want to add a proposal.
The unit of cost can be arbitrary, the obvious would be money, but it would also be possible to have a seat in a council as a resource.

Costs can be limited by a minimum amount, the default for this would be 0, as nothing speaks against a proposal that is free.
A higher minimum amount could be useful to eliminate minuscule proposals, where the administrative costs are higher than their implementation cost.
Proposals with a negative cost that bring in the used resource instead of spending it are imaginable but could have implications on the chosen scoring and selection algorithm and are omitted because of this.

Simultaneously to a lower end of the costs, there can be an upper limit on how much a proposal can cost. 
The default for this is the budget size, although, with negative costs proposals with a cost of greater than the budget are possible.
This is omitted for the same reason as above.
An upper limit can be used to forbid a monopoly proposal that uses up all the resources, or if the proposal is a candidate for a seat that obviously can not get more than one seat. 

A time can be set when no new proposals are allowed. 
This setting does not hinder discussion itself, so it can still be discussed existing proposals and arguments.
This is useful so that if the voting has not started yet, the proposals can be evaluated and edited.
And, which will be the usual case, it should not be allowed to make new proposals while the vote is held.

\section{Entering of Preferences}
At some point, the participants are able to vote for their preferred positions and rank them. 
Participants can change their vote as often as they please.
This can be possible parallel to when new proposals can be made or can start at a specific time.
Likewise, the ending of the voting time does not have to be set.
In this case, the results are shown directly and the votes can be changed indefinitely.

To vote for proposals they use approval voting, known from \autoref{sec:approval}, to express which proposals they want to see implemented.
The participant has no possibility to explicitly express a rejection as this would make the process more complicated. 
Furthermore, there is no reason to vote a proposal down as long as there is enough of the budget left to implement it. 
The majority will still win with their proposals.
An unwanted proposal will be repressed out of the budget if necessary.

Votes through approval have the advantage, that one can assume a safe default value for a missing vote.
Since a proposal can just be approved explicitly by a vote, there is no difference between a rejected proposal and a proposal a participant did not even notice or that did not exist at the time a participant made the last vote.

An original requirement was that proposals could be added after the vote has begun.\label{requirement-of-additional-votes}
This must not lead to participants having to cast their vote again or even modifying their vote without them knowing. 
Modifying proposals would lead to a more complex procedure, as the votes of the participants would have to be invalidated and contacted to vote again. 
This requirement was later made obsolete during the experiment but influenced some design decisions.

When two or more proposals have been approved the participant has the possibility to rank them by preference.
By default, proposals are ranked in the order in which they are approved, as this would be the natural behaviour the participant would expect.
The distance between two approved proposals is always the same. 
For the sake of simplicity, the participants do not have the possibility to express for example that two proposals are of the same worth for them.
Enabling this would have led the decision process to a utility-based one described in \autoref{subsec:utility}, which would have serious implications on the scoring algorithm.

While assessing the proposals the participant has the possibility of getting a quick look at arguments for or against.
These were made by other participants from the discussion before.
Just arguments one layer deep into the argumentation graph are shown, as the other participants were asked explicitly if they agree or disagree with a proposal.
This gives this layer the most expression of pro and contra arguments.
Moreover, a comprehensible presentation of a subgraph of the argumentation graph is a challenge on its own.
Participants are encouraged to jump right back into the argumentation with \gls{dbas} again if wished for. 

Contrary to earlier considerations no assumptions are made where the participant wants to enter into the argumentation.
Earlier in development, there was the consideration that if the participant already approved a proposal she would probably not like to argue against it.
This idea was rejected to enable uniform behaviour behind the interface.

The pro and contra arguments are randomized and truncated to three of each category to make the experience pleasant of discovering different aspects of a proposal by not too overwhelming the participant with information that is not asked for.
For participants who want to see all arguments, the possibility exists to list them all, as the goal is not to withhold any information that could influence the participant's choice.

\begin{figure}[ht]
    \centering
    \includegraphics[width=\textwidth]{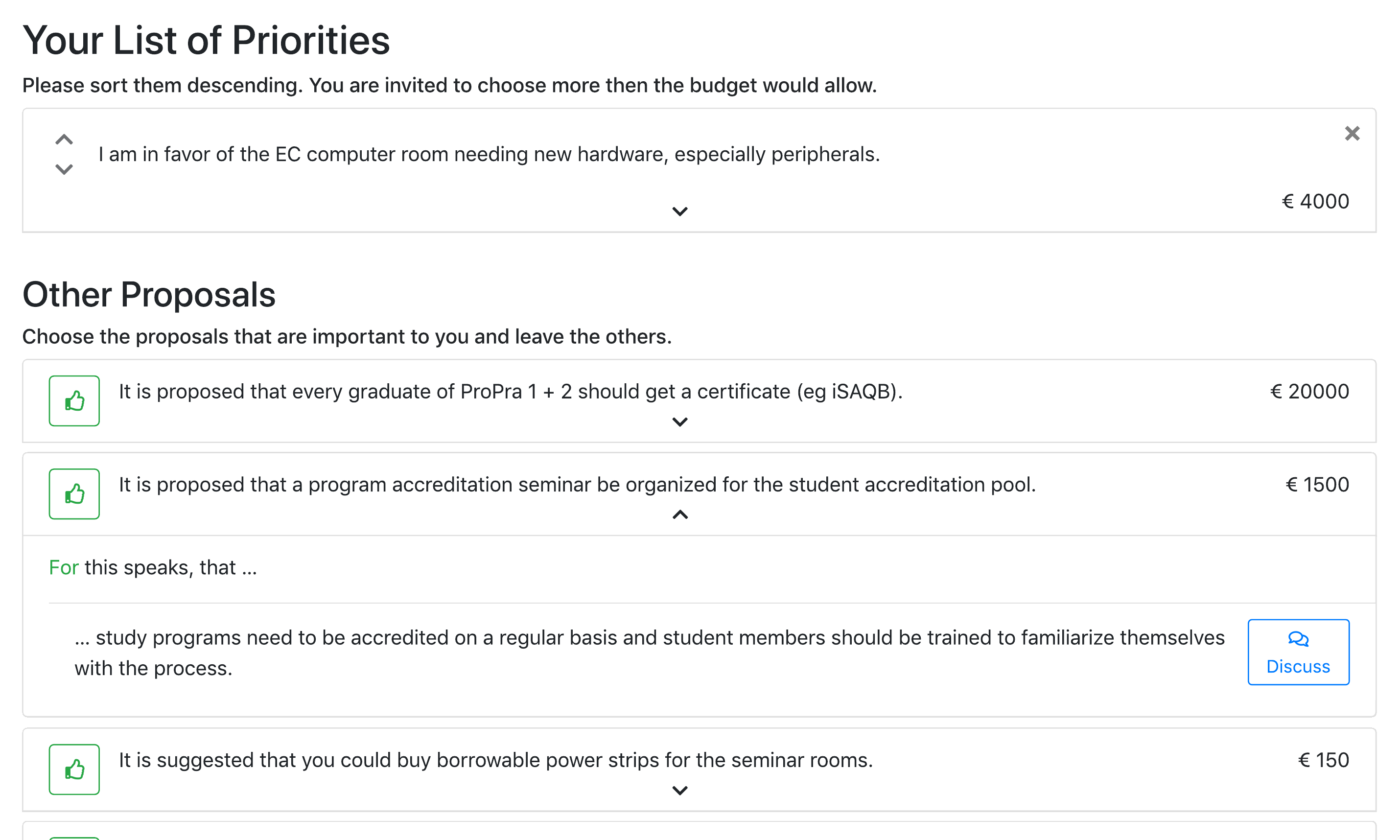}
    \caption[The voting interface]{The voting interface. Participants could approve proposals from below. Afterwards these proposals could be ranked to express priority. \\
    (This figure is translated from German).}\label{fig:voting-interface}
\end{figure}

\section{Determining the Winning Set}\label{sec:determine-winners}
After the participants have voted for and prioritized their favoured proposals, the result of winning proposals is determined.
Even though the result is available in real-time, we decided to hide the result until the voting phase was done.
We did this because we suspected that seeing the current result would animate participants to change their vote and check again on what had changed in the result. 
This would not be the desired behaviour as the votes of others would actually influence the own vote. 

For the result to be calculated the set of all preferences from the participants are gathered and scored with a mix of the approval and the Borda score explained in \Cref{chp:voting-and-scoring}.
The maximum number \(N\) as the highest score for Borda count is selected as the highest number of preferences a participant has chosen. 
With this, the Borda score for the preferences of each user is determined.
As before, the first preference gets the maximum score of \(N\), the second a score of \(N-1\), etc.
All proposals that were not preferred by the participant are getting the implicit, minimum score of \(0\).
The scores of every participant are summed up, just like with plain Borda score.

The behaviour of an implicit score for every proposal is needed because it could not be known beforehand how many proposals will be made.
It could well be so much of a burden to assess all the proposals that a participant has no interest in ranking them all, leading to the ranking of none of them. 

There would be other possible values for this truncated preference list, like the average of the scores the proposals would have received if they were appended to the end of the list.
This thought was abandoned to give a clear contrast between preferred and not preferred.

As a tiebreaker for this Borda score, the approval score ---1 if preferred, 0 if not--- was used, the ultimate tiebreaker is just the order in which the proposals were made. 
There was the thought to use the cost of a proposal as a tiebreaker because a proposal with fewer costs would allow for more winners in the end.
This seemed too much of an intervention into the vote to appear fair.

\section{Example}\label{sec:example}
A version of this example was given to the participants as a way to explain how the decision is made.

Suppose we have three participants Christian, Alexander and Markus.
We want to decide on three possible proposals. 
€ 10,000 are available as the budget.
The proposals could be: 
A hackathon with an estimated cost of € 4,000, a water cooler in the entrance area for € 2,000 and the modernization of a computer lab for € 7,000.

The participants are casting their preferences as shown in \autoref{preferences}.

\begin{table}[ht]
    \caption{Preferences of the three participants}\label{preferences}
    \begin{tabular*}{\textwidth}{l @{\extracolsep{\fill}} llll}
    \toprule
              & 1st priority (3 P) & 2nd priority (2 P) & 3rd priority (1 P) & rejected (0 P) \\ \midrule
    Christian & Water cooler       & Hackathon          & \multicolumn{1}{c}{-}                  & Computer lab   \\ \midrule 
    Alexander & Computer lab       & Hackathon          & Water cooler       & \multicolumn{1}{c}{-}              \\ \midrule
    Markus    & Hackathon          & Computer lab       & \multicolumn{1}{c}{-}                  & Water cooler   \\ \bottomrule

    \end{tabular*}
\end{table}

After the vote, their preferences get scored.
All their first preferences get the same points (P), the second ones get the same with one less than the firsts, etc. 
Alexander has submitted the most preferences with three and because of that, the score for all first preferences will be three points.
(Subsequently, all second preferences will get two points.)
Rejected preferences get no points in any case.

The sum of the points for the proposals are shown in \autoref{aggregated-points}:

\begin{table}[ht]
    \caption{Aggregation of Preferences}\label{aggregated-points}
    \begin{tabular*}{\textwidth}{l @{\extracolsep{\fill}} ll}
        \toprule
     & Points & Costs \\ \midrule
    Hackathon & 7 & € 4,000 \\ \midrule
    Computer lab & 5 & € 7,000 \\ \midrule
    Water cooler & 4 & € 2,000 \\
    \bottomrule
    \end{tabular*}
\end{table}

As the budget is just € 10,000 not all proposals can win at the same time.
Because the hackathon has the most points of them all, this proposal is included in the winning set, leaving € 6,000 for the remaining two proposals.

The proposal for the modernization of a computer lab will cost € 7,000. 
This is too much for the remaining budget.
The proposal can not fit in the budget without displacing the winner and so it will not be included in the winning set.

The water cooler has a cost of just € 2,000 that will fit in the € 6,000 of the remaining budget. 
Because of this, the proposal will be a winner, too, leaving € 4,000 of unspent budget. 

Proposals that nobody voted for ---which therefore got zero points--- will never win, which is an unlikely but possible event.
The expectation is, that the one making the proposal will also agree on it.

\subsubsection{Draw in Points}
Assume that a fourth participant, Martin, takes part.
With Martins preferences the tables were updated. (See \autoref{tab:preferences_4} and \autoref{tab:aggregated-points_4})

\begin{table}[ht]
    \caption{Updated preferences of the four participants}\label{tab:preferences_4}
    \begin{tabular*}{\textwidth}{l @{\extracolsep{\fill}} llll}
    \toprule
              & 1st priority (3 P) & 2nd priority (2 P) & 3rd priority (1 P) & rejected (0 P) \\ \midrule
    Christian & Water cooler       & Hackathon          & \multicolumn{1}{c}{-}                  & Computer lab   \\ \midrule 
    Alexander & Computer lab       & Hackathon          & Water cooler       & \multicolumn{1}{c}{-}              \\ \midrule
    Markus    & Hackathon          & Computer lab       & \multicolumn{1}{c}{-}                  & Water cooler   \\ \midrule
    Martin    & Computer lab       & Water cooler       & Hackathon
       & \multicolumn{1}{c}{-} \\
    \bottomrule
    \end{tabular*}
\end{table}

\begin{table}[ht]
    \caption{Updated aggregation of preferences}\label{tab:aggregated-points_4}
    \begin{tabular*}{\textwidth}{l @{\extracolsep{\fill}} lll}
        \toprule
     & Points & Costs & Approvals\\ \midrule
    Computer lab & 8 & € 7,000 & 3 \\ \midrule
    Hackathon & 8 & € 4,000 & 4\\ \midrule
    Water cooler & 6 & € 2,000 & 3 \\
    \bottomrule
    \end{tabular*}
\end{table}

Now the situation has changed. 
Because Martin prefers the computer lab and has sorted the hackathon to the third place, the points gained by both are equal.
The situation is resolved by taking into account the number of participants, who actually voted for a proposal. 
Independent on how they have sorted their preference.

Christian did not vote for the computer lab, while everyone approves to the hackathon. 
Because of that the computer lab just gets an approval score of three and the hackathon a score of four. 
The hackathon wins again, if only in the second instance.
The rest of the scenario plays out the same as in the example of three participants.

\section{Problems with the Estimation of Costs}\label{sec:cost-problems}
A problem for participants is that they can not reliably estimate the cost of a proposal. 
Projects in public space are often a lot more expensive than they are if they were done by a private person. 
This is because of labour costs, the need for certified equipment and several other influences that do not apply to private persons.

Often proposals can occur, that does not have a specific cost point. 
The participant would like to have a specific amount of funds for a proposal but would do fine if it would get less.
This observation leads to a conflict for the participant.
On the one hand, the participant would like to have the most amount of funds possible and would estimate the costs for the proposal as high as needed.
This would result in a lower chance of getting in the winning set if the proposal does not get the highest score in the vote.
On the other hand, the participant is therefore encouraged to make a cheaper estimate in order to receive any funds at all.

Coming back to the example in \Cref{sec:example}, Alexander estimated a cost of € 7,000 for the modernization of the computer lab.
A budget of € 4,000 was left of the budget. 
It did not match the estimated cost, but it would be imaginable that a computer lab could be improved with € 4,000 as well. 
Should Alexander have made a smaller estimate?
If the proposal with the water cooler had not been made, then there would be € 6,000 left. 
This would have left Alexander with less than he had hoped for.

This leads to the thought that participants could make proposals with more confidence if instead of a fixed price point they could make proposals with a range of cost. 
While this would be better for the participant making the proposal, it would be hard to let all participants vote on a proposal with a possible wild range of cost. 
Because there can be no guaranteed cost, this could seem deceptive.
The following section explains, why correcting the cost without evil intentions is still not recommended and could lead to trust issues in the procedure.

\subsubsection*{Corrections of Costs}
The costs of a proposal have to be fixed, as later modifications of them could influence the vote. 
The situation that someone had already voted for a proposal with certain costs and then drastically changed would not have been fair and would have changed their vote. 
This would be bad in either way and could lead to dissatisfaction.
If for example, participants had voted for a proposal with an estimated cost and the cost was drastically corrected upwards, then the participants may have never voted for the proposal in the first place, as it may have been too expensive for them. 
Subsequently, if the costs were corrected downward, it can not be assumed, that all participants who voted for this proposal, want it implemented if the implementation is going to be cheap. 
The participants could have quality concerns with a cheaper implementation. 
It has been noted that cost corrections would have to be made, but they can not take place if a participant could already vote.

\section{Possible Effects of Personal Incentives on the Discussion Climate}
By now, \gls{dbas} did not provide any outside incentive to show flag for a position, other than the own conviction.
Now, with the possibility of gain in the form of winning an election, be it for the win itself (remember: in \gls{dbas}, winning an argument does not exist) or a real-life personal benefit by getting funding for a desired project.
It can be assumed that participants argue differently if nothing is at stake for them.

It can be expected that arguments are made to weaken an opponents position, as attacking another argument could lead to more available funds for the own proposal. 
One could even agree with a position but decides to argue against it, because the position is indeed favourable, but may not be as favourable as another position. 
This form of argument was indeed added during the experiment.

By now there is no way to include positions into an argumentation other than as a conclusion to an argument, because of this \gls{dbas} does not have any way to represent this behaviour of users. 

\section{Voting Security}
In most kinds of votes, there are some security and privacy considerations to be made. 
These considerations have to be done to protect the participants and the process itself.

As the voting is conducted as a secret vote, no other participant is allowed to see what another participant has voted for, and it could even be argued that it can not be allowed to know if a specific participant has participated in the process or not.
The latter may be leaked, because \gls{decide} uses \gls{dbas} as its authentication backend.
\Gls{dbas} creates a user account for everyone who logged in at least once and these user accounts are publicly visible and most of the users who voted did not have an account in \gls{dbas} before.
Although the existence of an account is not proof that a user has voted, this is a strong indication.
As the participation in services from \gls{dbas} was not high while the vote was ongoing no new users other then from the vote were added.
This is a flaw that has to be taken into consideration in possible future processes.

As the participants were informed and had to agree that their submissions may be used anonymously for scientific purposes, these requirements had to be loosened for obvious reasons.
At least the person in charge of the procedure will have to have access to the votes, be it anonymously, to determine how many have participated in the voting process without attending the argumentation actively.

\subsubsection{Authorization and Authentication}
Unlike in other participatory processes, it was possible to ensure, that every participant can just vote once because participants have to log in through the \gls{idm} of the university.
In environments where there are not such \glsplural{idm} this can be a serious challenge. 

Outside of an institution like a university, other forms of authentication would have to be used. 
Registration and login through a common e-mail based system could lead to fake accounts generated by harmful participants.
This is something that has to be avoided, but it could lead to other problems, especially for emerging processes. 

E-mail authorization is known by almost all users of the internet and is, therefore, a low barrier to participate in a process.
Other means of authorization and authentication, like a state identification card number, such as the German eID card\footnote{https://www.bsi.bund.de/EN/Topics/ElectrIDDocuments/German-eID/german-eID\_node.html}, can be a bigger barrier.
This could result in lesser participation overall, threatening the existence of participatory budgeting, and e-participation processes as a whole.

Which forms of authorization and authentication could be used has to be decided on a per-process basis. 
For example, it could be possible to invite citizens of a neighbourhood to a participation process where a letter with a security code is sufficient, as the impact on a decision is contained and may not be as critical as a political vote.

Better security measurements have to be implemented if one considers deciding in a procedure where the stakes and therefore the incentive for manipulation is higher.

\section{Implementation Details}
The software developed for this experiment ---\gls{decide}\footnote{https://github.com/hhucn/decidotron}, to fit with the other \gls{dbas} extension \emph{discuss}\footnote{https://github.com/hhucn/discuss}--- is meant as a piggy back extension to \gls{dbas}. 
The workings of \gls{dbas} are not modified radically that would restrict other discussions.
Only the possibility to add an estimated price point to a proposal was added, but just for discussions that required this feature.
This allows running several discussions next to each other in a single instance.
Even parallel decision processes are possible. 

The changes on \gls{dbas} allow to set all timings in the database itself, allowing for automated switching between phases.
Since the timings are optional, it can be customized how different phases in a procedure play out.
For example in the experiment for this work, there was the spontaneous need to disable the addition of further proposals while existing proposals were reviewed.
It would also be possible, though, that results are always visible and the voting phase never ends.
This could be useful when proposals were made continuously by participants while winning proposals were processed at the other end by the administration and afterwards removed from the system\footnote{Refer to the process in Reykjavík in \cref{sec:iceland}}. 

Authorization and authentication were also coupled to the \gls{dbas} main system and was not handled separately, allowing for the matching between how the participants participated in the argumentation and how was voted. (See \cref{sec:collected-data})

\Gls{decide} is open source and under the MIT licence. It can be found on \url{https://github.com/hhucn/decidotron}.
\Gls{dbas} can be found under \url{https://github.com/hhucn/dbas} and was initially developed by Tobias Krauthoff as part of his doctoral thesis\cite{krauthoff_phd}.
\Gls{dbas} is now under ongoing development and maintenance by the chair for computer networks and communication systems at the Heinrich-Heine-University.

\chapter{Experiment}
To try out how participants make use of \gls{dbas} and \gls{decide}, an experiment was conducted with students of the computer science course at the Heinrich-Heine-University.

This implied that a fully controlled and observed scientific environment could not be created for the experiment, as the procedure was a true process of real money distribution.
The participants were told that the process was scientifically accompanied and their submissions will be used anonymously for this and future scientific works. 

If the aftermath of this experiment is satisfying for each party, there is the thought of using the same procedure in other courses. 
This would have the desired side effect that the procedures can be compared and evaluated next to each other.

The raw anonymous data (in German) can be found in the \hyperref[app:raw]{Appendix}.
The argumentation can be found online on \url{https://dbas.cs.uni-duesseldorf.de/discuss/verteilung-von-qualitatsverbesserungsmitteln}.
The decision system (now only showing the results) can be found here: \url{https://decide.dbas.cs.uni-duesseldorf.de/preferences/verteilung-von-qualitatsverbesserungsmitteln}. 
Both systems are subject to change.

\section{Setup}
For the setup of the experiment, the \gls{we} (wissenschaftliche Einrichtung / WE) of computer science provided € 20,000 of so called \gls{qvm} (quality enhancement funds).
These funds can be requested for the improvement of teaching in the course of computer science. 
Because the funds were not used to the full extent in recent years, this money piled up, with no idea on how to spend it.
It is assumed that most of the students do not even know about these funds or its purpose, so the decision was made to use the money for proposals the students provided, giving them the control and maybe advertising that such things are possible.

For this, certain restrictions had to be made:

\begin{enumerate}
    \item\label{only-students} Only properly enrolled students of computer science may bring in proposals and vote. 
    \item A proposal may not cost more than € 20,000, for obvious reasons. 
    \item A proposal may not cost less than € 100, as with a minuscule estimate the administrative cost would outnumber the cost of the proposal itself. The value was estimated and there was no restriction by the \gls{we}.
    \item\label{prohibited-proposals} A proposal has to be made in the scope of the \gls{qvm}, that means it has to make an improvement to the teaching and it must be able to be implemented by computer science department without entering into the area of responsibility of other departments. To be explicit, no proposals for changes to the infrastructure of the university can be accepted. 
\end{enumerate}

These rules were published on the launch page and were accessible through a prominent button in the interface where new proposals could be made. 
Examples of proposals that can not be implemented as they do not follow Rule~\ref{prohibited-proposals} were visible there.

Rule~\ref{only-students} was enforced through the LDAP directory of the university, which is publicly accessible for every student and employee.
Participants could just login with their id and password that they use for other services in the university.
The LDAP directory allows a filter to just accept students who enrolled in computer science.
While the trust in the correctness of this directory was not absolute, the directory was the single best way to authenticate students. 

The process started on the 23. April 2019 and was initially planned to last for two weeks.
Divided up in two phases, each one week in duration, the first phase would be focused on the submission and argumentation of proposals.
The second phase would just open the voting system.
This was done, so that participants will not just take part at the beginning of the procedure, when this time just a few proposals would already be made, and vote on them, missing the later proposals.

Shortly after the start of the procedure, the discovery was made, that a third phase has to be introduced, stretching the experiment to nearly three weeks. 
All students were invited to the process by e-mail over a common e-mail distribution list.
There was no specific target group, as this would be against the spirit of the \gls{qvm}.

\begin{figure}[hp]
    \centering
    \includegraphics[width=\textwidth]{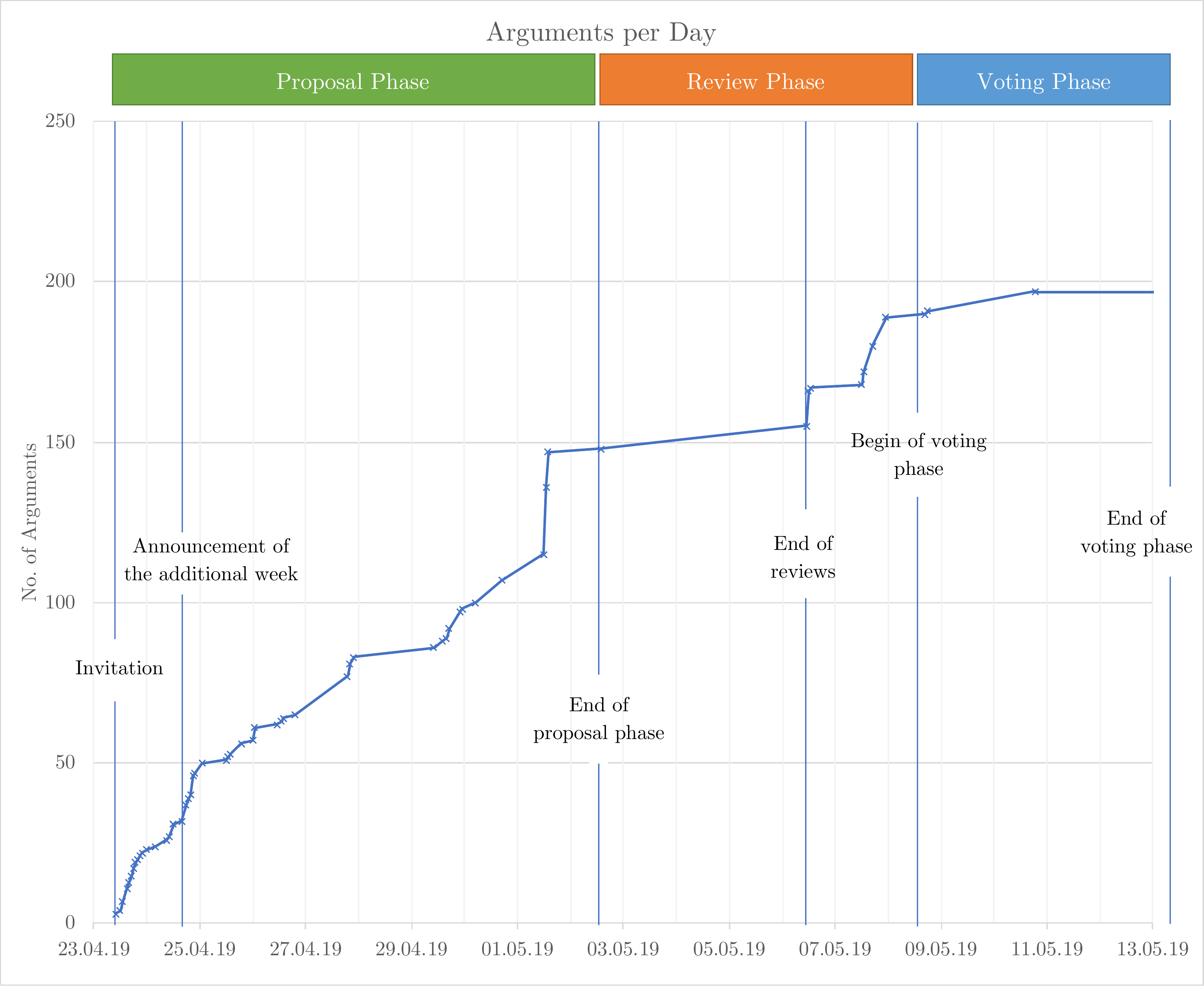}
    \caption[Timeline of the experiment]{This is the timeline of the procedure. The participation process started on the 23rd of April 2019 at 12:36 and ended on the 13th of May 2019 at 12:00. 
    Each vertical line represents the moment an e-mail was sent to the participants to inform them about the ongoing procedure.
    The number of arguments is measured once every hour and marked by a cross.
    }\label{fig:timeline}
\end{figure}

\section{Phase 1 --- Proposals and Argumentation}
For the experiment, the live, public version of \gls{dbas} was used, like in the field experiment conducted in 2017\cite{Krauthoff2017a}, where everyone from the computer science department was asked to argue about how to improve the course of computer science. 
As expected the most participation was on the day of the notification, with a degrading amount of participation the following days. 

All in all 52 registered students participated in the argumentation.
Eleven of these had used \gls{dbas} before ---not necessarily in the first field experiment--- and 36 registered for the first phase. 
The remaining five participants joined the argumentation later in the process in phase 3. Contrary to the first field experiment there was no data collected about participants that did not login, refer to \cref{sec:collected-data} for detailed information.

\begin{figure}
    \centering
    \includegraphics[width=.5\textwidth]{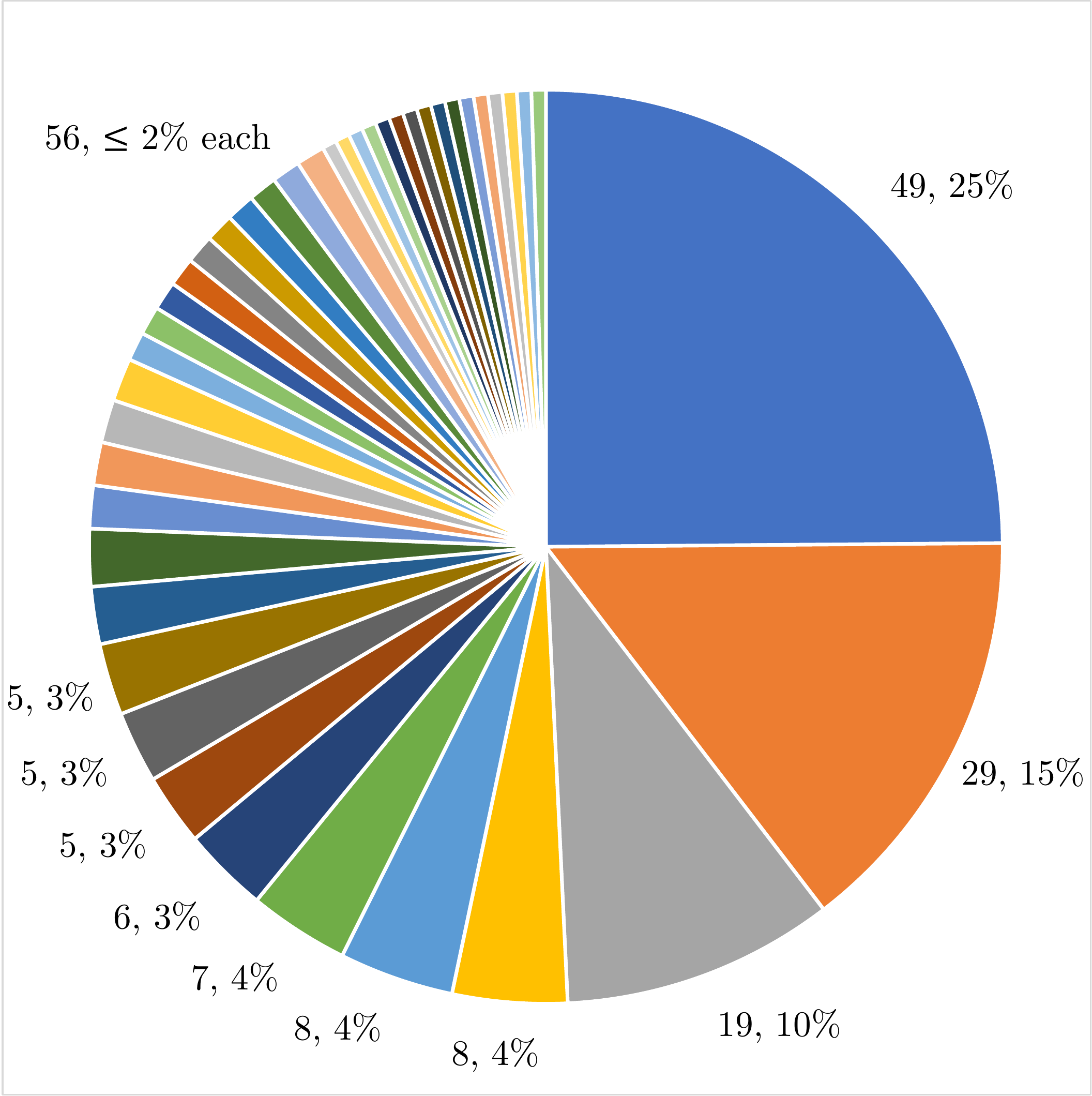}
    \caption[Proportion of participants of the total number of arguments.]{Proportion of participants of the total number of arguments. The full circle represents 197 arguments.}
    \label{fig:participation-by-user}
\end{figure}

\Cref{fig:participation-by-user} shows the share each user has in the total number of arguments. 
Even more than in the first field experiment with \gls{dbas} the argumentation was driven by a few participants. 
In the case of the user with the highest share, the arguments were mostly not added throughout the course of the argumentation but in a relatively short time.
This is the cause for the steep growth of the graph in \cref{fig:timeline} at the 01.05.2019 12:00 mark.

\begin{figure}
    \centering
    \includegraphics[width=\textwidth]{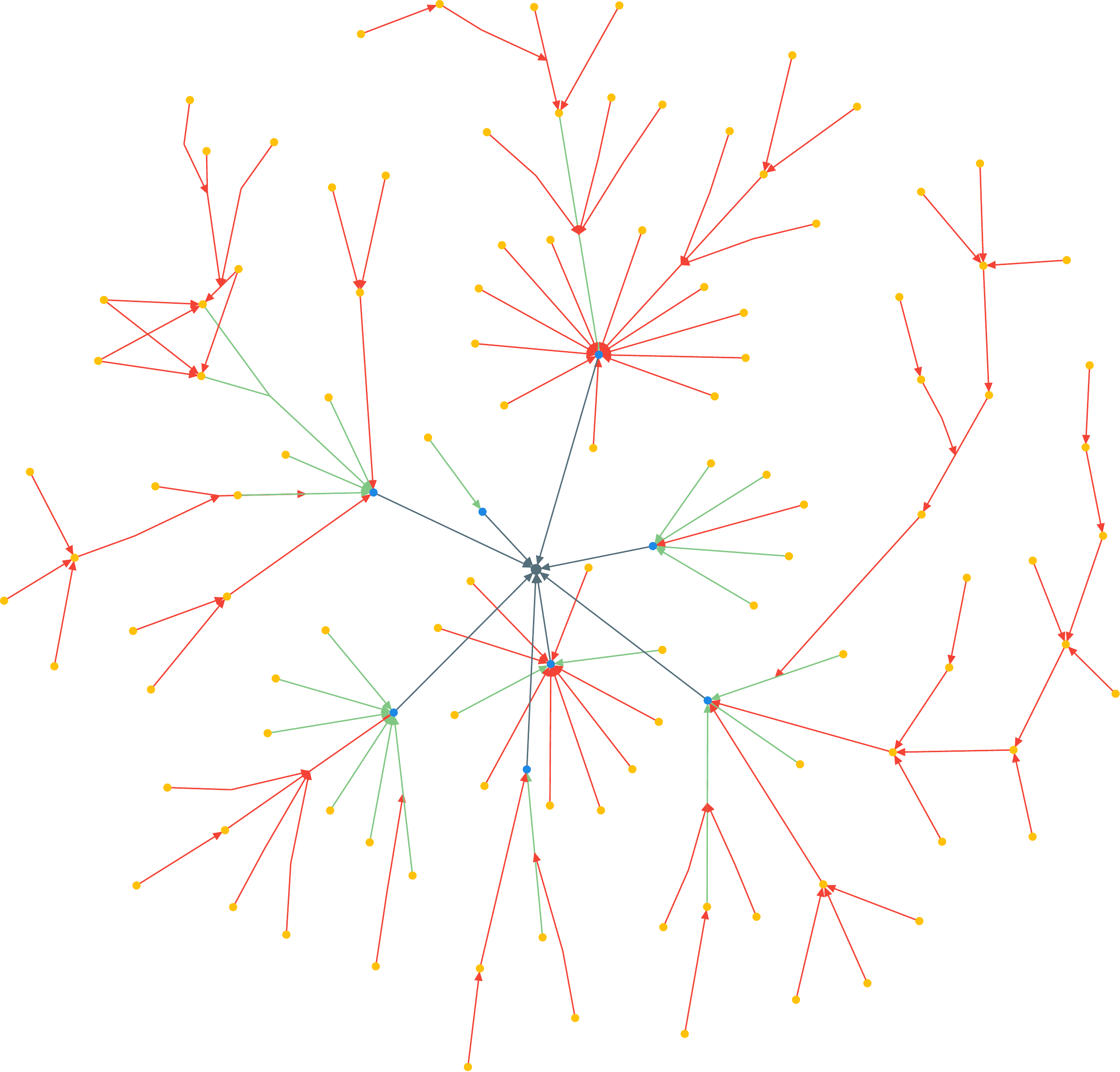}
    \caption[The final argumentation graph]{The final argumentation graph, after the filtering of phase 2. The grey point is the \emph{issue}, blue points are proposals (\emph{position}s), yellow points are \emph{statement}s. Green and red arrows represent \emph{arguments} pro or contra the argument, they are pointing at.}
    \label{fig:graph}
\end{figure}

\subsubsection*{The Need to have a Review of the First Phase}
After the first day, it was found that participants did not read or straight ignored the rules the proposals had to adhere to. 
Several proposals were made that were against Rule~\ref{prohibited-proposals} and even mentioned in the examples.

Furthermore, it was found that participants did not stick with \gls{dbas}'s format of statements and that they made richer statements contradictory to the idea of indivisible statements in \gls{dbas}. 
This suggests that either the interface of \gls{dbas} is not clear enough on how to use the tool, or that there is the need to carry out one's proposal further. 
The interface does not allow for further information, or differently, is not designed for this.

Both of the mentioned issues forced the addition of an unplanned third phase after this phase, in that these proposals were filtered according to Rule~\ref{prohibited-proposals} and clarified, so that in the last phase, the voting phase, each proposal would be unmistakable.


\section{Phase 2 --- Correction of Proposals}\label{subsubsec:phase2}
Initially, the plan for the experiment was that there are just two phases.
There should not be any influences other than editorial changes.
The second phase would consist of further argumentation.
This was abandoned, as it became clear that most of the proposals were made in a way that would make them impossible to be included in the vote.

Reasons for this were:
\begin{enumerate}
    \label{enum:reasons-for-failure}
    \item Proposals did not comply with the rules.
    \item Proposals were underspecified. 
    No clear intent was visible.
    \item Proposals were specializations of another proposal.
\end{enumerate}

Several proposals were clearly not implementable and were strict against the published rules.
This has gone so far that proposals were made that were explicitly mentioned in the rules, as not implementable. 
For example, it was mentioned that the computer science department could not improve the insufficient Wi-Fi coverage, as this is within the discretion of the Center for Information and Media Technology (ZIM). 
Nevertheless, this was one of the first proposals, arguing that with the amount of money it should be possible to override that limit.

Another reason was that proposals were made broadly, meaning they were ambiguous, with an unknown scope of action needed. 
The intention of the author was not clear for the other participants as well as for the person who later has to implement them.
Example of this would be the proposal to use the funds to generate more funds.
The proposal did not mention on how exactly this was imagined, whether it was intended to use the money to arrange a fair to attract donating companies or for advertising of the course. 
An additional minor example would be, that a participant proposed to provide a seminar room with power strips, this had to be specified to say that the power strips had to be certified for e.g.\ fire resistance and that they can not be permanently in the room, but can only be borrowed.
The proposals that were not clear in their intent were suspected to harm the later voting process.
They could lead to a misunderstanding with the participants and could have led to a later dissatisfaction if the proposals won the voting process, but were implemented in a radically different way as someone had imagined.

A third week was added to the whole process to reflect the proposals and make editorial changes, for that the possibility to add proposals after the first phase had to be disabled.
It was expected that there will be some proposals which can not be implemented, but after the first day of the process, it became visible that this problem was bigger than originally expected.
Like mentioned in \cref{requirement-of-additional-votes}, before the start of the experiment the plan was that proposals could be made in the voting phase as well, which dictated a lot of the design decisions for the system.

\section{Phase 3 --- Voting}\label{subsubsec:phase3}
The last phase began with eight out of 21 possible proposals, on which the participants could vote on. 
The proposals had editorial changes, which do not alter the original intention. 
In one case three proposals were merged into a single one.
To ensure transparency, a notification was sent out to inform the participants on why some proposals had to be removed or changed. 

Because of the delays introduced by the extra phase, this phase was shortened from the planned week to five days, ending on a Monday.
Even though nearly three times the number of active participants took part in this last phase (142 voters), the interest in discussing the existing proposals was low, as seen in \cref{fig:timeline}.
Just ten arguments were added to \gls{dbas} in this last week, contrary to expectations that further proposals will be made and discussed at this stage.

\begin{figure}[h]
    \centering
    \includegraphics[width=.7\textwidth]{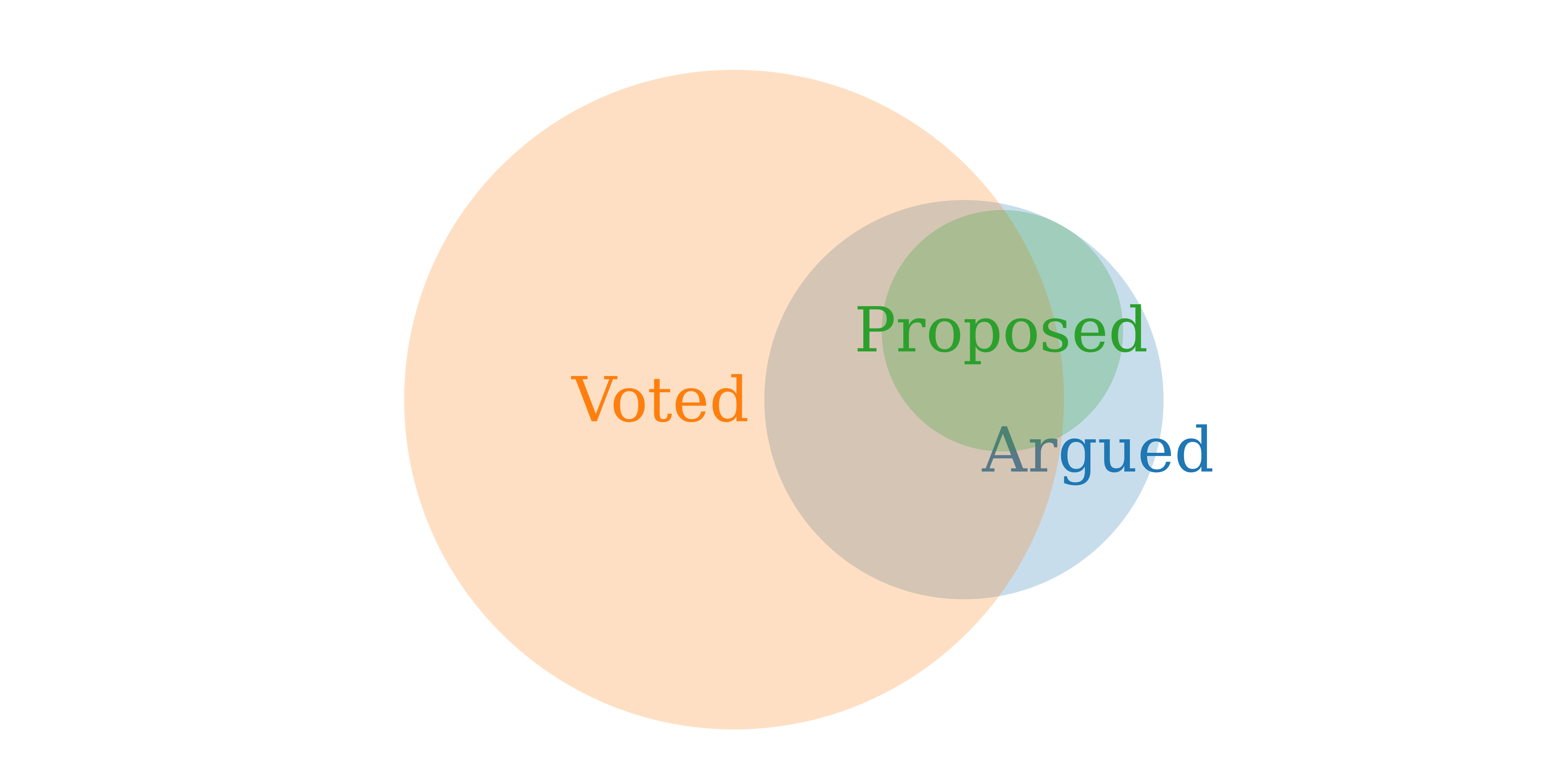}
    \caption[Proportion of participants in all levels of participation.]{Proportion of participants in all levels of participation. Proposed has to be a subset of Argued.
    \\Voted: 142, Argued: 52, Proposed: 19, 
    \\Argued \& Voted: 39, Proposed \& Voted: 14}
    \label{fig:venn-participation}
\end{figure}

It should be noted that from the 52 participants in the argumentation 18 of them did not cast a vote. 
Five of them were authors of proposals themselves, with four of their proposals rejected.
As seen in \cref{fig:venn-participation} this is a noticeable amount of missing votes. 
This is contrary to expectations, as an interest in the argumentation process would indicate interest in the whole process where the final vote would mark the most important and only way to influence the made decision directly.

\section{Results}
\begin{table}
    \centering
    \includegraphics[width=\textwidth]{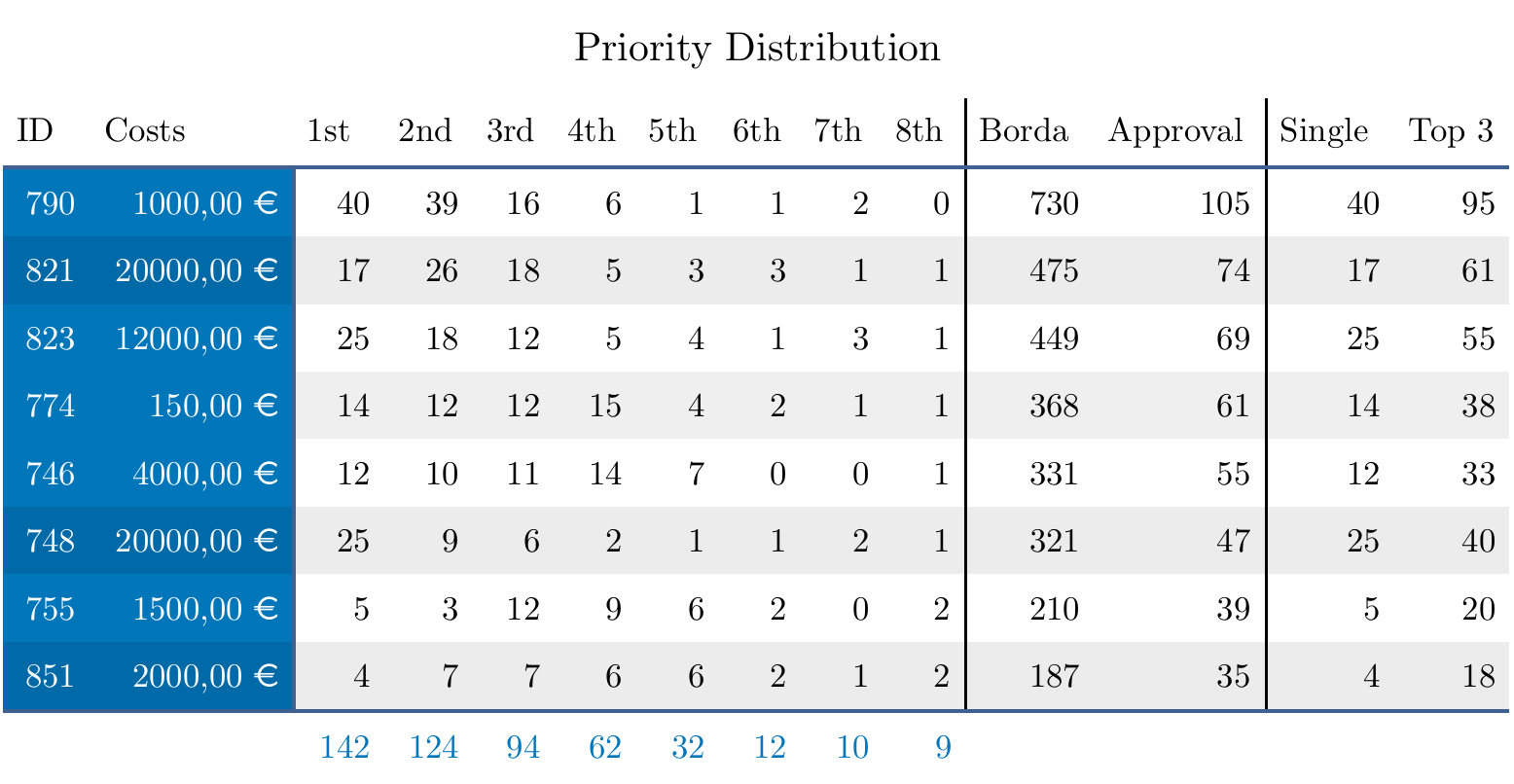}
    \caption[Distribution of votes for proposals]{Distribution of proposals (ID) that were chosen with a specific priority. 
    Accompanied by the used Borda and approval scores, as well as simulated scores for a vote with just a single vote and a Top 3 priority vote.
    Sorted first by Borda score. The tainted proposals lost in the process.}\label{tab:preferences}
\end{table}

The results were published twenty days after the start of the experiment. 
They were available exactly at the end of the experiment, as the calculation could be done in real-time, as mentioned in \cref{sec:determine-winners}.

A total of 142 participants entered their preferences, outweighing the numbers of participants in the discussion immensely. 
Whether this is a result of missing information about participants who were not logged into \gls{dbas} or if the burden of arguing is just higher than casting a vote cannot be answered. 

Notable is that there was nowhere near as much participation in the discussion itself, like at the beginning of the procedure as can be seen in \cref{fig:timeline}. 
This may be the result that the discussion itself stagnated or that there is not an immediately perceivable reason in further argumentation. 

Of the remaining eight proposals, five were elected to be a winner.
All in all € 18,650 of the available € 20,000 will be utilized. 
Of the three proposals that were not voted, there were two proposals costing € 20,000.
The expectation was indeed that either one of the proposals with a cost of € 20,000 will win or (nearly) everything else.
One proposal (821; `Certification courses such as LPIC, MCSA or CCNA should be offered.') got the second most points, but was too expensive with € 20,000. 
As the first proposal (790; `More programming language courses should be offered.') just costs € 1000 it could be argued that with just a small reduction of 821s cost, the result could be completely different.

The results in Table\ref{tab:preferences} show, that the outcome of the vote would be the same if just the approval score would be used to rank the proposals.
This shows that in the participants in the voting process are mostly aligned in their desires.
A different result for ranking by Borda or approval score would mean that either, (1) there is a small group of voters who just voted for a proposal that the majority does not approve or that (2) there is a proposal that is approved by a majority but only as a lesser preference, while the top preferences are varied a lot. (1) would result in a high Borda score but in a low approval score. (2) would lead to a low borda score, with a high approval score. 

\begin{figure}[htbp]
    \centering
    \includegraphics[width=\textwidth]{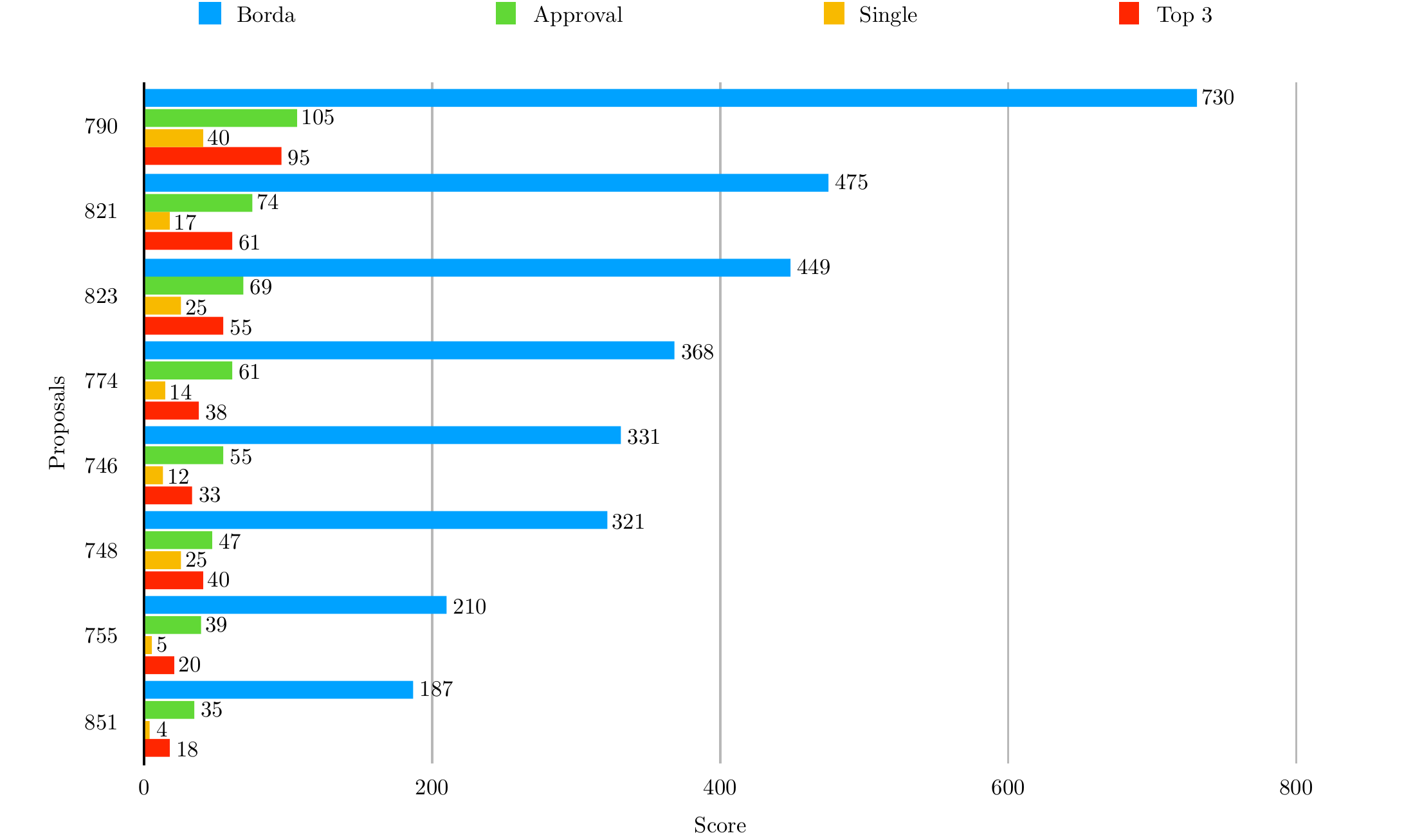}
    \caption{Score distribution of the proposals.}\label{fig:scores}
\end{figure}

\subsubsection{Results of other Scoring Methods}
Although just Borda and approval scores are used for the result, simulations of other scoring methods were done to investigate if the result would be different.
The scores can be found in \cref{tab:preferences} and \cref{fig:scores}.

The two other scores chosen are: 
\begin{description}
    \item[Single-Vote] that simulates the case where every participant can just vote for one proposal. 
    The proposal gets a score of 1. 
    This system is included because of the frequency this voting system gets used in day to day life.

    \item[Top 3 Approval] is the truncation of the approval score to the top three proposals. 
    It is not unusual in voting procedures to enforce a restriction like this\footnote{Refer to the process in Wuppertal mentioned in \cref{sec:wuppertal}}.
\end{description}

Actually, an approval score of the Top 2 preferences would yield another result. 
As seen in \cref{tab:preferences}, proposal 755 gains a lot of approval score with its votes as the third preference. 
Without this additional score, proposal 851 would have won.

Another usual restriction would be that participants could just approve to proposals in such a way that the approved proposals could not break the budget.
This has a hit on the expensive proposals, as with a cost of € 20,000 they could just be included in the preferences as the first proposal. 

\begin{figure}[htbp]
    \centering
    \includegraphics[width=.76\textwidth]{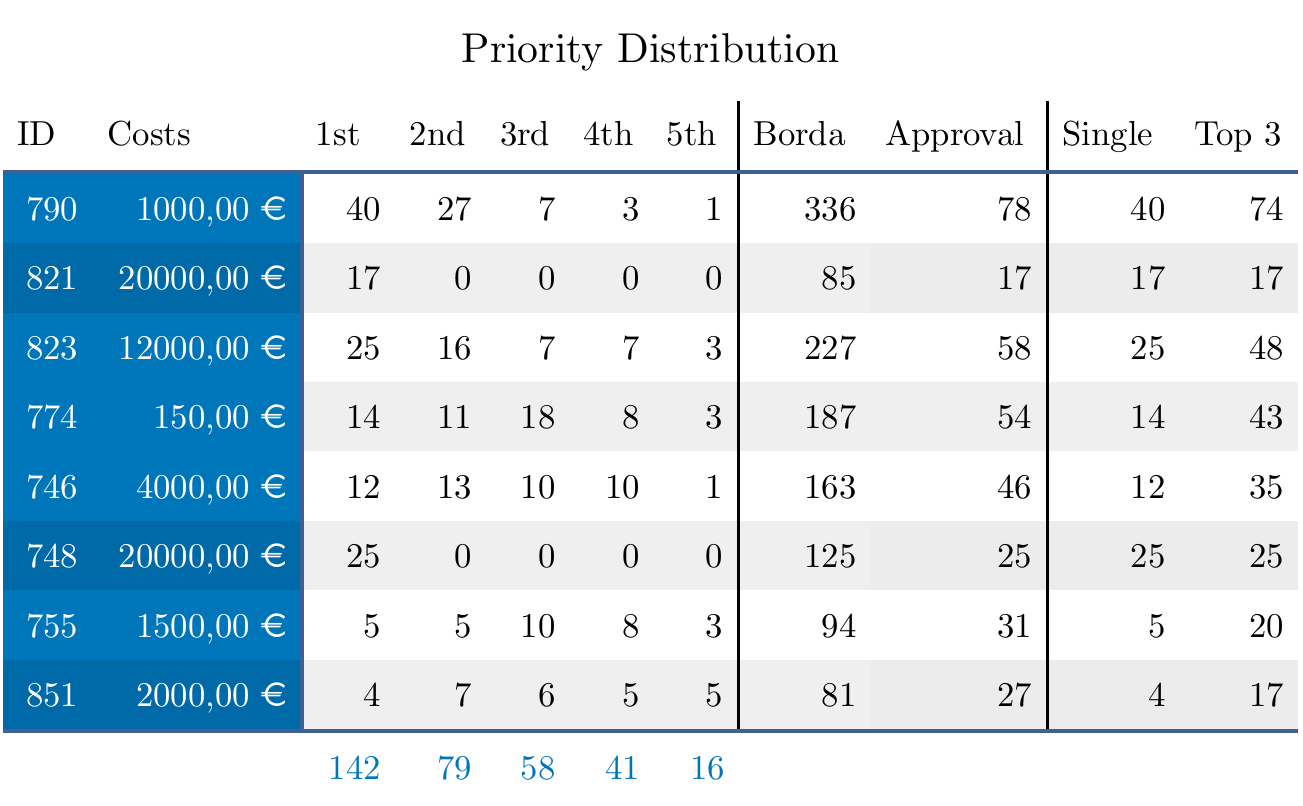}
    \caption{Priority distribution, if the participants were restricted by the budget}
    \label{tab:preferences-restricted}
\end{figure}

Although the outcome does not change, the budget filling € 20,000 proposals are punished severely. 
It has to be decided from procedure to procedure if this is desired. 
On the one hand, the knowledge about the general interest in these proposals may be lost, as they are not chosen any more.
On the other hand, it can serve as an educational tool.
The goal of this would be to teach citizens about financial dilemmas resulting from a restricted budget. 
See \cref{tab:preferences-restricted}.

\section{Feedback from the survey}\label{sec:survey-feedback}
\begin{figure}[p]
    \centering
    \includegraphics[width=\textwidth]{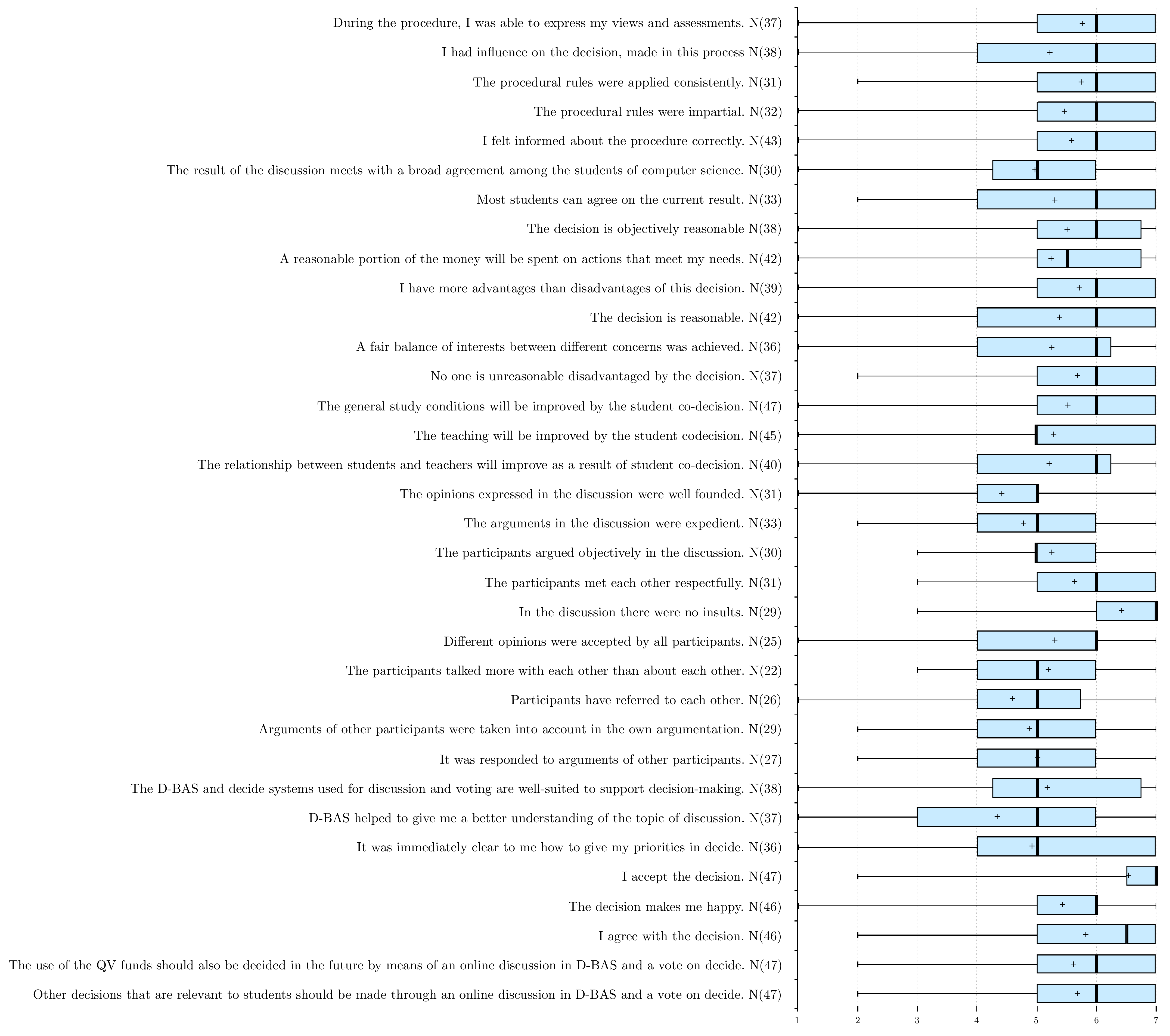}
    \caption[Survey result boxplot]{Questions and their results. 1 is equal to \emph{not correct at all} and 7 is \emph{In any case, correct}. The boxes are showing the range of the median 50\% of scores, while the whiskers are showing the minimum and maximum scores. The thick bar in the diagram denotes the median. The \(+\) is the average.}
    \label{fig:survey-boxplot}
\end{figure}

With the announcement of the winners every computer science student was invited to a survey, here no difference was made, whether a student just argued, voted, did both or none of it.
This survey was done in cooperation with the Department for Social Sciences from the Heinrich-Heine-University.
A cropped list of results is shown in \cref{fig:survey-boxplot}.
A number of questions are left out, as they are not relevant for this work, but for the social science department. 
The full results will be published in the near future.

All in all the procedure was received positively by the participants.
The weakest results were at the question \emph{D-BAS helped to give me a better understanding of the topic of discussion}. 
As seen in \cref{fig:survey-boxplot}, more than a fourth of the survey participants answered towards \emph{not correct at all} since there is not a more specific result, this can mean that either for some participants \gls{dbas} is too complicated or that the participants had trouble to keep track of the whole topic. 
This is a problem not yet tackled in \gls{dbas}.

The results for the question \emph{it was immediately clear to me how to give my priorities in \gls{decide}} hints that improvements on the interface of \gls{decide} are needed. 
It seems to polarize the participants in the survey more than expected, and while a considerable amount of participants saw it as practical, it has to be remembered that the procedure has to be inclusive and should be unambiguously understandable by everyone.

Pleasant is the results for \emph{Most students can agree on the current result}, \emph{No one is unreasonably disadvantaged by the decision} and their like.
They show that the procedure is considered fair and testify to the general satisfaction of the participants with the outcome, which is clearly an objective when trying to enable participation.

The by far best results are for the last five questions (except for the question about insults in the argumentation), all of them were positive to exceptional positive. 
This validates that the done experiment was useful and shows that more participation possibilities are desired.

\section{Collected Data}\label{sec:collected-data}
\gls{dbas} collects which argument was made by whom. 
This data is available to the public to everyone who participated in the process, as the default setting is that the real name of the participant is shown, to generate an environment in which everyone gets the encouragement to make a serious contribution to the process instead of hiding behind their anonymity.
Nevertheless, the participants had the choice to choose that their name was not publicly visible.

Since the amendments to the \gls{gdpr}, \gls{dbas} no longer collects data on unregistered users, because of this it is unknown how many people visited the argumentation without logging in. 
A login was just required when adding arguments to the argumentation or interact otherwise with \gls{dbas}, like participating in editorial changes to reported arguments.

\Gls{decide} only records the votes of a participant and that a participant voted at all. 
The exact data of how the participants voted is of interest in the validation and later evaluation of the aggregation algorithm itself and is, of course, kept anonymous.
The votes in \gls{decide} are authorized by \gls{dbas} as the backend.
Because of this, it is possible to investigate if participants who voted also participated in the argumentation itself.

\section{Future Improvements to D-BAS and decide}
Although the process was successful, there are a number of improvement points if the process should be repeated in the future. 
These are features that were out of the scope of this work or \emph{nice to have} features.
This would include tighter integration between \gls{decide} and \gls{dbas}.
By now the systems are on separate sites with different UIs, therefore the jumping between the voting UI and the argumentation is kind of unintuitive.

No data on whether participants were more satisfied with the simpler overview of \gls{decide} is available.
It is clear that the list of \gls{decide} overview is more suited to get an overview of the larger realm of arguments than to navigate each node in the argument graph via the means of a dialog.
It should be evaluated what kind of UI improvements in \gls{dbas} ---and subsequently, \gls{decide}--- are fitting to improve understanding of the discussion.

A number of more complex improvements, that should be thought of that were hindering in the experiment. 
They could greatly improve the experience and possibilities of the approach in this work.

\subsection{Loosen the Constrictions on Statements}
In the experiment, it was found that most of the times participants do not stick to the given grammatical structure of `I think proposal A is right because of \dots'.
This is of great concern, as the correct grammatical form is needed to later build meaningful sentences, with which the participant is confronted.
Three common violations against the form imposed by \gls{dbas} were identified.

\begin{enumerate}
    \item\label{violation_simple} Simple grammatical violations. \\
            Example: \emph{Buy a 3d printer.} \\
            To fit into the form, this should be formulated like this: I think \emph{we should buy a 3d printer}
    \item\label{violation_multiple} Two or more arguments done in a single argument. \\
            Example: \emph{Because there are already computers [\dots]. Most students use their own devices anyway.} \\
            This would have to be split into an argument which is supported by the second argument.
    \item\label{violation_additions} Statements which cannot be modeled as an argument. \\
            Example: \emph{We should set up coffee machines. Like they are in the building of business administration.} \\
            The information, that there is an example for the argument does not fit into the argument graph, as the sentence is neither a support nor an attack against anything, just plain information.
\end{enumerate} 

Violation~\ref{violation_simple} can be fixed by the moderation system provided by \gls{dbas}, someone has to report this.
Violations like this indicate that the participant did not understand what the task was. 
This could be due to a bad interface. 

Violation~\ref{violation_multiple} could origin in a too slow interaction flow through \gls{dbas}.
The participant may not be aware that there are ways to structure arguments like that.
It would be advantageous if the participant directly has the option to create a subtree of arguments when entering the argument.

The last Violation~\ref{violation_additions} could be tackled by allowing participants to enter free text. 
This must be limited to the bare minimum, as+ it could undermine the whole purpose of structured argumentation if used incorrectly, but has the potential to enable a better understanding in other participants which have to deal with this argument.

All in all, it seems like \gls{dbas} is misunderstood and subsequently misused by the participants in some cases, because \gls{dbas} is an unknown form of arguing, with constraints which may overwhelm the perception of a new participant.
Often there is no possibility to formulate a complex argument with proves, references and so on, simply.

\subsection{Extending Proposals}
In the experiment, there was a proposal made to buy a 3d printer for € 1,000. 
A short time later, a participant realized that printing material is needed for the operation of a 3d printer and so another proposal was made to buy printer material for € 1,337. 
This proposal makes sense if the first proposal is accepted, but does not when the 3d printer was rejected. 
Because of that the two proposals should be related, or to be more specific, the first proposal should be enhanced with the idea of the second proposal. 
This could be implemented by allowing a proposal to have extensions as well, which act like proposals themselves. 
In a case like this, the participants later have to vote not just on the first proposal itself, but which extension they want to see implemented alongside the original proposal. 

The extension system could accommodate a variety of proposed changes. 
Extensions could be proposals to build upon the initial idea, by extending it with funds, but also to tone down the proposal. 
This could be the case, when a participant has the opinion that a proposal is good, but may be too ambitious, so the participant adds a limitation to the proposal. 
A proposal could be `We should buy a cheaper 3d printer. This would cost € 300  less.' thus reducing the costs of the proposal from € 1,000 to € 700 when this extension is voted for. 

This raises the question on how to score the proposals to get the result of the vote. 
Each combination of proposal and selectable extension could be combined to a single possible winner with the restriction that just one of the combinations could win at a time. 
The participant would have to be given the same prioritization option as in the choice for the proposal itself. 
This could prove challenging from the user experience point of view, as the way of expressing one's preference cloud become too complex.
\chapter{Participatory Budgeting Implementations}\label{chp:budgeting_implementations}

This chapter presents some forms of participatory budgeting. 
It has to be noted, that there are various processes out there, and each of them is different.
Most of the times, these procedures are suited for their specific situation, as most of them are done with a political background or other intervening forces.

This overview is limited to relevant examples, containing the beginnings of participatory budgeting, an example on how to combine offline and online participation and an fairly similar and hugely successful procedure.

\section{Porto Alegre (Brazil) since 1989}

The first large scale implementation of participatory budgeting began in 1989 in Porto Alegre, a city in southern Brazil with a population of more than 1 million\footnote{census from the 1st of September 1991}.
This started due to the lack of trust in the local government from the population\cite{wampler2000}.
It began with less than a thousand participants and grew to an estimated number of 40,000 in 1999\cite{bhatnagar} and continued to be an annual participation process, but stumbled in recent years due to political changes in the city\cite{Abers2016}.

\subsection{Process}
The process in Porto Alegre starts at the beginning of each year, with the preparation of the whole process. 
In the first half of the year, the government informs about the finances and determines regions in which the city is divided up.
Each of these regions is assigned a \emph{Quality of Life Index}, which is later used for resource distribution. 
On the other side, the participants analyse this information and gather followers to form interest groups. 
The participation takes place in offline meetings where the local population comes together.
In these meetings, first proposals from the region are presented and discussed, together with proposals which the government plans to implement.
At the end of this phase, a rough selection of possible projects is made.

In the next phase, the government publishes estimates of cost for the proposals, that can be implemented and were made in the previous phase.
The proposals are discussed in more detail and then votes are held for them. 
Each region votes by themselves for the priority of projects they would like to have implemented. These votes take place in the regular meetings, in which two representatives are chosen to represent the region in a \emph{Municipal Budget Council} which has the responsibility to decide which projects will be implemented.

\subsection{Comparison}
While this participatory process is strictly offline, this process is included in this work because the process is considered to be the origin of participatory budgeting, which has led to various procedures in Brazil and worldwide.

Deliberation takes places between informal factions in their corresponding regions.
This is more advisable when conduction an offline participation process in a large geographic area where the matters of decisions are local most of the time. 
Also, there is still not a direct influence of the people, as they vote for a representative that makes the final decision.
This is a necessity when there is no (digital) infrastructure to enable a large group of people to participate directly.

Notable is the distribution of financial resources.
As the process is done in different regions with a proportion of the whole budget, several processes are done in parallel.
Through the \emph{Quality of Life Index}, the budget is not distributed evenly.
According to the needs of each region, with poorer, less developed regions are getting a larger percentage of the budget while wealthier, more developed ones getting less. 
This sounds like a great idea, but can be hard to implement without an exact metric of \emph{need} for participating areas.
\section{Wuppertal 2017}\label{sec:wuppertal}

In 2017 the city of Wuppertal, Germany conducted a participatory budget process with around 3324 participants.
The goal was to allocate a fixed budget of € 150,000.
The process was therefore not just a consulting process, in which a council decided about what has to be deployed, but a somewhat more direct method of participation.

The process itself was done with the EMPATIA platform, which was developed by the \mbox{EMPATIA}\cite{empatia17} project\footnote{https://empatia-project.eu/}.
A project by a consortium of European information and communication technology partners, one of which, Zebralog\footnote{https://www.zebralog.de}, had the responsibility for the process.
The participation took place in multiple phases, wherein every phase the citizens had a different way of participating. 

There was a final report which describes and summarizes the whole process.\cite{empatia18_final_report}.

\subsection*{Phase 1 --- Collection of ideas}

In the first phase of the participation process, the citizens could submit new ideas how to spend parts of the budget. 
The submissions were possible by a number of methods to allow a broad spectrum of citizens to participate.
These include an online form, e-mail, postal, phone but were also possible in person at one of many street stands.

Along with the collection of ideas and afterwards participants had the possibility to rate and comment on the ideas to aid the implementing organizations in reviewing the submissions. 
For the last week, it was not possible to submit ideas anymore, but only voting was allowed.
Of the 267 submitted ideas, 109 (a rough \emph{Top100}) were chosen by the administration to advance to the next phase. 
The remaining 158 were filtered out because they did not fit with the criteria of the process. 

\subsection*{Phase 2 --- Review by the citizens}
The Top100 ideas were presented at an open event for the citizens, a week after the first phase ended. 
In a workshop where participants and representatives of the municipality could discuss the proposals together and decide which ones are going to be the finalists, on which the votes are held upon.
These final ideas were checked again by the administration and 32 were approved to be possible to be implemented.
These were compiled into a \emph{Top30} list and a further 16 proposals were excluded due to their cost and the missing ability of the city to implement them. 

\subsection*{Phase 3 --- Voting}
The voting on the \emph{Top30} proposals lasted three weeks. 
Participants had the chance to cast up to five votes either online, in person in the town hall or at a special \emph{voting party} held at the beginning of the voting phase.
It was not possible to vote for a single proposal multiple times and the votes had equal weight.
The votes were summed up and the top-ranked proposals were added to the result set if their cost fitted into the budget.
If they did not fit, they were simply excluded from the winning set.
An exception to this rule was that the administration reconsidered the costs of the proposal which was going to lose with around € 13,000 of the proposed € 20,000 missing.
They concluded that the proposal could still be implemented with the remaining € 7,000, even if that meant that the proposal had to be cut somewhat.

For the participants to cast their vote online, they would have to be verified to prevent electoral fraud. 
This was realized through SMS and landline telephone verification. 
The results of the vote were not shown while the vote was still in progress, even though the result was calculated in real time like it is possible in our software. 
This was possible because it was a direct decision, not a form of the consulting process. 
Out of 5,325 votes from 1,627 participants, 4,761 were conducted online.

\subsection*{Running Process in 2019}
In 2019 the city of Wuppertal started a new participatory budgeting process\footnote{https://talbeteiligung.de/topic/buergerbudget} with a budget of € 165,000. 
This fits as two years are the budget period of Wuppertal, so it indicates that there was imminent interest in a continuation of this kind of decision making. 
The similar amount of allocated budget could indicate that the interest of the municipality is not to outsource their decisions to the people, but to show the participants that they can provide significant input to local politics. 

The process is no longer run by EMPATIA and Zebralog, but instead, the process was taken over by {{wer|denkt|was}}\footnote{https://werdenktwas.de/}. 
Nonetheless, the new process is like the first one.

\subsection{Comparison}
This process is similar to the approach in this work, in that it enables a direct decision making by the participants. 
In comparison to other cities in Germany, this is one of the first processes, which allows the participant to make the final decision\cite{empatia18_final_report}, while the government can just limit the available choices.
Participatory budgeting processes were mostly done in a consulting manner, obscuring how the final decision was made by leaving it to the usual way of a ruling council. 
This way a participant does not have proof that her involvement was significant to the outcome of the process.

Problematic in the light of the experiment done in this work is, that a stretched out procedure with several phases requires most of the participants to stay active with the matter. 
A luxury that can not be expected everywhere, particular when the processes are small an not well marketed.

\subsubsection{Differences}
A difference to our process is, that we do not allow for so-called multi-channel participation,
where the process is not strictly available online, but allowed for parallel participation via different media, after which results are merged in a single online system. 
Although this is a good way to establish initial interest, this could prove more expensive when the scale of this kind of process grows.

\begin{enumerate}
    \item Proposition phase
    \begin{enumerate}
        \item A Comment section like in \cref{sec:comment-section} is used.
        \item Proposals have an upper limit of costs.
        \item Proposals can be done either online or offline in various forms, including in person, by mail or by telephone.
    \end{enumerate}
    
    \item Review Phase. 
    \begin{enumerate}
        \item Review takes place more than once with different granularities.
        Starting with a rough review from the municipality, over an assessment by the citizens and at last a feasibility analysis. 
        The last to reviews are done offline and together with representatives of the organizer.
    \end{enumerate}

    \item Voting Phase
    \begin{enumerate}
        \item Votes are limited to five elements.
        \item There is no prioritization.
    \end{enumerate}
\end{enumerate}

\section{Open Active Voting (Iceland)}\label{sec:iceland}
Open Active Voting is a tool developed by Citizens Foundation\footnote{https://citizens.is} in Iceland.
It is open source and available under the GNU AGPL v3 license\footnote{https://github.com/CitizensFoundation/open-active-voting}.
The tool aims to enable participants to allocate a fixed budget for the neighbourhood to proposals with a known cost.
It is used in the annual participatory budgeting process\cite{neighbourhood} in Reykjavík, Iceland, which started in 2011.
The budget of every year is 450 million Icelandic króna (about 3.3 million €\footnote{As of 24.04.2019: 1 króna is equal to € 0,0074})

The proposals are submitted by the participants before the voting process.
During the election in spring 2010, 40\% of the population of Reykjavík visited the proposal page (at this time, the electoral system was)\cite{your-priorities-iceland}. 
After a month of collecting ideas, the municipality evaluates the ideas, whether they are feasible or not and estimates a cost for each one to be implemented.

The ideas are bound to one of ten neighbourhoods in Reykjavík and each participant can just cast his vote for one of them.
Each neighbourhood has a different budget, based on the population density.
The neighbourhoods are not assigned through or checked against any residents register. 

Votes are cast by selecting the desired set of proposals that fit into the budget assigned to the neighbourhood. Just proposals from a single neighbourhood can be chosen, although participants are not prohibited to change their neighbourhood and voting differently. The last vote counts.
Each proposal that was voted for gets a weight of one.
Additionally, a single proposal can be marked as a favourite, doubling its weight. 
At the end of the voting process, all weights are summed up.
The winners are chosen separately for each neighbourhood and are chosen in the same way as presented in this work, by gobbling up the top-rated proposals and skipping them if they do not fit in the remaining budget.

The procedure does not have an end, instead, the council of Reykjavík promised to evaluate the Top10 ideas each month, deciding whether the proposals are getting implemented. By this means 64\% of all proposals were accepted.

\subsubsection{Security}
Notable about this process is, that there are different verification systems in place for different phases of the process.
Proposals can be submitted with an unverified login by e-mail or even a Facebook account, lowering the entry level for participation. 
This allows for a deeper investment of the participants in the process, which will likely increase the chances that the participant returns to the system again to vote.

The voting itself is secured by either a text message verification or the Icelandic government's electronic identity system \emph{IceKey}\footnote{https://www.island.is/en/icekey-e---certificate/about-icekey/}. 
According to official statistics the e-identity system is in use by approximate 270,864 private persons\footnote{https://www.island.is/en/indentification-services/login-service-statistics/}, at an overall population of Iceland of around 340,097 as of 2019.

\subsection{Comparison}
The voting system itself has similarities to the approach in this work.
It allows participants to vote for as many proposals as they like, with the difference that the sum of the chosen proposals cannot exceed the budget, an idea we rejected as we did not see any gains from this limitation.
It actually hides the interest of participants in the proposals that are excluded because of their cost but teaches the voters about civic budget restrictions\cite{presentation_citizens_foundation}.

The proposals in this procedure are supported by pro and contra arguments that can be debated.
The arguments can be voted up and down by the participants.
These pro and contra debates resemble the way \gls{dbas} is used through the voting phase to display arguments and allows further participation.

All in all this procedure has a lot of similarities to the one used in this work.
The major difference is, that the costs of proposals are not set by the participants, but rather by a ruling authority that can estimate the costs precisely, eliminating the problems mentioned in \cref{sec:cost-problems}.

\subsubsection{Differences}
\begin{enumerate}
    \item Proposition phase
    \begin{enumerate}
        \item A comment section like in \cref{sec:comment-section} instead of an argument graph is used. 
        This section is separated in a pro and a contra list, which lessens the gap to the system used in this work.
        \item Costs for the proposals are not set by the participants. 
        This is a critical difference from this work.
        \item Proposals are specific to an area, but as the participants can choose which area they feel they belong to and the budget cycle for this does not result in different processes.
    \end{enumerate}
    
    \item Proposals are reviewed. 
    \begin{enumerate}
        \item Costs are added by the municipality. 
    \end{enumerate}

    \item Voting
    \begin{enumerate}
        \item Votes from a single voter cannot exceed the budget.
        \item There is no end to the voting process. The budget gets updated every year and the leftover proposals are recycled in the procedure.
        \item There is no prioritization.
    \end{enumerate}
\end{enumerate}

A demo system is in place to explore the interface and its possible functions: 
\url{https://ktest.betrireykjavik.is}.

\chapter{Conclusion}
In this work, it has been explored, if a decision process restricted by funds and costs and supported by an argumentation system can be successful and if it will lead to acceptance by the participants. 
As the experiment has shown, this is possible and the participants are satisfied with the procedure and the results. 
It has been found that even with strict rules, a decision-making process is not sufficiently resilient against inadequate proposals. 
The addition of the reviewing phase resulted in a hard separation of the argumentation and the voting process. 
This has cut the procedure practically in half, as the voting phase actually became just a traditional vote.

This leads either to the path of a supervised procedure, or the approach of \gls{dbas} to structure data as it enters the system. 
In this case, the latter seems to be an inadvisable solution because, unlike previous applications of \gls{dbas}, the information for a proposal must be validated against an external source.
The supervision of a specialized group, like the municipality, is advised in these kinds of decisions.
Future procedures in the style of this experiment are advised to provide means to easily access and modify the argumentation, like merging arguments or removing proposals.
Currently, there is no way for untrained users of \gls{dbas} to do such kind of modifications.

\subsubsection*{Future Work}
In order to test the acceptance of the participants, the experiment should be repeated in such a way that there are not so many winners.
It can be argued, that the satisfaction in the outcome is high because most of the proposals were able to be included in the winning set of proposals and there were no highly controversial proposals.

Moreover, the results of this work do not mean that an unmonitored decision-making process in a large society is not possible. 
If the goal is to make a less significant decision that does not dependent on resources that are outside the control of the deciding system, the argumentation and voting could happen without supervision.

If it turns out, that participants have great trust in a deciding system, it could be explored, how a decision can be made without a traditional vote, just by anticipating the behaviour of the participant in the argumentation. 
A possible scenario would be, that in the way \gls{dbas} interacts in a dialog with the participant, the system could ask directly for opinions about positions, which hold knowledge for a decision.
Further advancement of this would appear in the example of the scenario of the demo discussion in \gls{dbas}, where a family discusses whether they should get a cat or a dog, without any system to make a decision.
An observing, deciding agent could make a decision for the family, without the direct involvement of the members in any kind of vote.
This could happen just by analysing the argument's relation to the family members.
However, this would require that participants actually participate in some form in the discussion and it can be seen in the results of the experiment, that the majority of participants did not actively participate in the argumentation process itself, but rather just vote. 
This could just be the result of less hassle for the participant.

\appendix

\chapter{Raw Data}\label{app:raw}

\begin{table}[h]
    \centering
    \resizebox{\textwidth}{!}{%
    \begin{tabular}{lll}
    ID & Proposal & Cost \\
    790 & mehr Programmiersprachekurse angeboten werden. & € 1000 \\
    821 & Zertifizierungskurse wie z.B. LPIC, MCSA oder CCNA angeboten werden sollen. & € 20000 \\
    823 & wir mit der Unterstützung des ZIMs einen Hackerspace inkl. 3D Drucker einrichten. & € 12000 \\
    774 & man für die Seminarräume ausleihbare Mehrfachsteckdosen anschaffen könnte. & € 150 \\
    746 & der Computerraum im EG neue Hardware, insbesondere Peripherie, braucht. & € 4000 \\
    748 & jeder Absolvent von ProPra 1+2 ein Zertifikat bekommen sollte (bspw. iSAQB). & € 20000 \\
    755 & ein Programmakkreditierungsseminar des studentischen Akkreditierungspools ausgerichtet wird. & € 1500 \\
    851 & das Abgabesystem (AUAS) eine Auto-Speichern Funktion braucht. & € 2000
    \end{tabular}%
    }
    \caption[The raw proposals and their costs in German.]{The raw proposals and their costs in German. They were prefixed in the UI to a full sentence.}
    \label{tab:raw-proposals}
    \end{table}

\iftrue
\clearpage
\begin{longtable}[h]{llllllll}
    \caption[Raw preferences of proposals from 1st to 8th preference.]{Raw preferences of proposals from 1st to 8th preference. The row order is random.}
    \label{tab:raw-preferences}\\
    1st & 2nd & 3rd & 4th & 5th & 6th & 7th & 8th \\
    \endfirsthead
    \multicolumn{8}{c}%
    {{\bfseries Table \thetable\ continued from previous page}} \\
    1st & 2nd & 3rd & 4th & 5th & 6th & 7th & 8th \\
    \endhead
    821 & 790 &     &     &     &     &     &     \\
    790 &     &     &     &     &     &     &     \\
    774 & 790 & 755 & 746 &     &     &     &     \\
    790 & 748 &     &     &     &     &     &     \\
    821 &     &     &     &     &     &     &     \\
    790 & 748 & 823 &     &     &     &     &     \\
    748 & 790 & 821 & 823 &     &     &     &     \\
    823 & 790 & 746 & 755 &     &     &     &     \\
    748 & 823 & 774 & 790 & 821 &     &     &     \\
    823 & 851 & 790 & 774 &     &     &     &     \\
    790 & 823 & 774 & 746 &     &     &     &     \\
    748 & 821 &     &     &     &     &     &     \\
    851 & 790 & 746 &     &     &     &     &     \\
    790 & 746 & 823 & 774 & 755 &     &     &     \\
    774 & 823 & 755 & 790 & 821 &     &     &     \\
    823 & 790 & 821 & 774 & 755 & 851 & 748 & 746 \\
    821 & 790 & 755 & 748 &     &     &     &     \\
    790 & 821 & 823 &     &     &     &     &     \\
    790 & 746 & 823 &     &     &     &     &     \\
    790 & 821 &     &     &     &     &     &     \\
    774 & 790 & 746 & 755 &     &     &     &     \\
    790 & 823 &     &     &     &     &     &     \\
    748 & 821 & 790 &     &     &     &     &     \\
    823 &     &     &     &     &     &     &     \\
    790 & 774 & 823 & 821 &     &     &     &     \\
    823 & 851 &     &     &     &     &     &     \\
    748 & 821 & 823 &     &     &     &     &     \\
    746 & 851 & 748 & 821 & 774 & 823 & 790 & 755 \\
    746 & 851 & 790 & 755 & 748 & 774 & 821 & 823 \\
    746 & 790 & 821 & 774 &     &     &     &     \\
    821 & 748 & 790 & 755 & 851 &     &     &     \\
    823 &     &     &     &     &     &     &     \\
    851 & 755 & 746 & 790 &     &     &     &     \\
    851 & 823 & 821 & 746 & 774 &     &     &     \\
    755 & 823 & 774 & 851 &     &     &     &     \\
    790 & 774 & 755 & 746 & 851 &     &     &     \\
    748 &     &     &     &     &     &     &     \\
    790 & 851 & 746 & 823 & 774 &     &     &     \\
    746 & 821 & 755 & 790 & 823 &     &     &     \\
    774 & 821 &     &     &     &     &     &     \\
    790 & 755 & 851 & 746 &     &     &     &     \\
    746 & 821 &     &     &     &     &     &     \\
    748 &     &     &     &     &     &     &     \\
    823 & 790 & 755 & 746 &     &     &     &     \\
    748 & 821 & 790 &     &     &     &     &     \\
    790 & 774 &     &     &     &     &     &     \\
    790 & 823 & 821 & 746 &     &     &     &     \\
    790 & 746 & 755 & 774 & 821 &     &     &     \\
    748 & 823 & 746 &     &     &     &     &     \\
    821 & 748 & 790 & 755 & 746 &     &     &     \\
    821 & 790 &     &     &     &     &     &     \\
    821 & 823 & 748 & 774 &     &     &     &     \\
    823 & 746 & 790 & 774 &     &     &     &     \\
    790 & 821 & 755 &     &     &     &     &     \\
    774 & 790 &     &     &     &     &     &     \\
    774 & 790 & 821 & 851 &     &     &     &     \\
    746 & 823 & 851 & 774 &     &     &     &     \\
    748 & 790 & 774 & 746 &     &     &     &     \\
    790 & 746 & 774 & 821 & 755 & 851 & 823 & 748 \\
    823 & 774 & 755 & 790 &     &     &     &     \\
    755 & 790 & 823 &     &     &     &     &     \\
    821 & 790 &     &     &     &     &     &     \\
    774 & 821 &     &     &     &     &     &     \\
    748 & 790 & 823 & 774 & 746 &     &     &     \\
    746 &     &     &     &     &     &     &     \\
    790 & 821 & 748 & 755 & 746 &     &     &     \\
    823 & 790 & 774 &     &     &     &     &     \\
    748 & 790 & 821 &     &     &     &     &     \\
    823 & 851 &     &     &     &     &     &     \\
    774 &     &     &     &     &     &     &     \\
    774 & 790 & 755 & 821 &     &     &     &     \\
    790 & 746 & 821 & 774 & 755 & 748 & 823 & 851 \\
    755 & 774 & 851 & 746 &     &     &     &     \\
    790 & 821 &     &     &     &     &     &     \\
    790 & 748 & 746 & 755 & 823 & 821 & 774 & 851 \\
    821 & 790 & 774 &     &     &     &     &     \\
    790 & 823 & 851 &     &     &     &     &     \\
    790 &     &     &     &     &     &     &     \\
    823 & 790 & 774 &     &     &     &     &     \\
    746 &     &     &     &     &     &     &     \\
    823 & 746 & 821 & 774 & 851 & 755 &     &     \\
    790 & 821 & 746 &     &     &     &     &     \\
    748 & 790 & 821 &     &     &     &     &     \\
    821 & 790 &     &     &     &     &     &     \\
    790 & 823 & 774 & 746 &     &     &     &     \\
    823 & 790 & 774 &     &     &     &     &     \\
    823 & 790 & 821 & 748 & 774 &     &     &     \\
    748 & 821 & 790 &     &     &     &     &     \\
    790 &     &     &     &     &     &     &     \\
    748 & 823 & 851 &     &     &     &     &     \\
    748 & 821 & 746 & 823 &     &     &     &     \\
    821 & 748 &     &     &     &     &     &     \\
    755 & 746 & 774 & 823 & 851 & 790 & 748 & 821 \\
    790 & 823 & 774 & 746 & 755 &     &     &     \\
    748 &     &     &     &     &     &     &     \\
    755 & 790 & 823 & 774 &     &     &     &     \\
    821 & 774 & 755 & 851 & 790 &     &     &     \\
    746 & 774 &     &     &     &     &     &     \\
    821 &     &     &     &     &     &     &     \\
    748 &     &     &     &     &     &     &     \\
    774 & 746 & 790 & 755 & 823 & 821 & 851 &     \\
    746 & 790 & 821 &     &     &     &     &     \\
    790 & 746 &     &     &     &     &     &     \\
    821 & 790 &     &     &     &     &     &     \\
    823 & 821 & 851 &     &     &     &     &     \\
    821 &     &     &     &     &     &     &     \\
    748 & 821 & 823 & 746 & 851 & 755 & 790 & 774 \\
    823 & 821 & 790 & 774 & 755 &     &     &     \\
    790 & 823 &     &     &     &     &     &     \\
    823 & 748 & 790 &     &     &     &     &     \\
    790 & 821 & 746 & 774 & 823 &     &     &     \\
    790 & 823 & 821 & 774 & 746 &     &     &     \\
    790 & 823 & 851 &     &     &     &     &     \\
    748 & 790 & 746 &     &     &     &     &     \\
    823 & 790 & 748 & 746 &     &     &     &     \\
    821 & 851 &     &     &     &     &     &     \\
    790 & 774 & 823 & 746 &     &     &     &     \\
    790 & 821 & 748 &     &     &     &     &     \\
    823 & 790 &     &     &     &     &     &     \\
    748 &     &     &     &     &     &     &     \\
    790 & 821 &     &     &     &     &     &     \\
    748 & 821 & 790 &     &     &     &     &     \\
    823 & 790 & 821 & 755 &     &     &     &     \\
    748 & 821 & 790 &     &     &     &     &     \\
    851 & 790 & 748 & 821 & 746 & 774 & 823 & 755 \\
    746 & 774 & 823 & 851 &     &     &     &     \\
    748 & 755 & 821 & 790 & 746 &     &     &     \\
    748 & 821 & 790 &     &     &     &     &     \\
    790 & 823 &     &     &     &     &     &     \\
    790 & 774 &     &     &     &     &     &     \\
    774 & 790 &     &     &     &     &     &     \\
    774 & 790 &     &     &     &     &     &     \\
    774 & 790 & 755 & 851 & 746 &     &     &     \\
    790 &     &     &     &     &     &     &     \\
    823 & 774 & 790 & 851 &     &     &     &     \\
    823 & 821 &     &     &     &     &     &     \\
    823 & 790 & 821 &     &     &     &     &     \\
    823 & 748 & 821 &     &     &     &     &     \\
    774 & 748 & 821 &     &     &     &     &     \\
    746 & 774 & 790 & 823 & 851 & 821 &     &     \\
    790 &     &     &     &     &     &     &     \\
    821 & 790 &     &     &     &     &     &    
    \end{longtable}
\fi

\backmatter
\bibliographystyle{static/alphadin} 

\bibliography{bib/master}

\printglossary[]
\printglossary[type=\acronymtype]

\printindex

\begin{otherlanguage}{ngerman}

\chapter*{Ehrenwörtliche Erklärung}

Hiermit versichere ich, die vorliegende \thesistypegerman{} selbstständig verfasst und keine anderen als die angegebenen Quellen und Hilfsmittel benutzt zu haben.
Alle Stellen, die aus den Quellen entnommen wurden, sind als solche kenntlich gemacht worden.
Diese Arbeit hat in gleicher oder ähnlicher Form noch keiner Prüfungsbehörde vorgelegen.

\vspace{3cm}

\noindent Düsseldorf, \thesissubmissionday{}. \DTMmonthname{\thesissubmissionmonth} \thesissubmissionyear{} \hfill \thesisauthor{}

\end{otherlanguage}

\end{document}